\documentclass[12pt]{article}
\usepackage{amssymb}
\usepackage[dvips]{graphicx}
\usepackage{amsmath,amsfonts,bm,amssymb,dsfont}

\usepackage{xcolor}

\usepackage{expl3}
\ExplSyntaxOn
\cs_new_eq:NN \Repeat \prg_replicate:nn
\ExplSyntaxOff

\def\l{\left}
\def\r{\right}

\def\ql{\textquotedblleft}

\def\1o2{{1\over2}}

\def\a{\alpha}
\def\b{\beta}

\def\d{\delta}
\def\D{\Delta}
\def\m{\mu}
\def\n{\nu}
\def\k{\kappa}

\def\La{\Lambda}
\def\s{\sigma}

\newcommand{\fig}[1]{Fig. \ref{#1}}

\newcommand{\mc}[1]{\mathcal{#1}}
\newcommand{\tx}[1]{\text{#1}}
\newcommand{\be}{\begin{equation}}
\newcommand{\ee}{\end{equation}}
\newcommand{\bea}{\begin{eqnarray}}
\newcommand{\eea}{\end{eqnarray}}
\newcommand{\bi}{\begin{itemize}}
\newcommand{\ei}{\end{itemize}}
\newcommand{\bc}{\begin{center}}
\newcommand{\ec}{\end{center}}

\begin{document}

\begin{titlepage}
\vfill
\begin{flushright}
ACFI-T15-04
\end{flushright}

\vfill
\begin{center}
\baselineskip=16pt
{\Large\bf Symmetry Breaking Vacua in Lovelock Gravity}
\vskip 0.15in
\vskip 10.mm
{\large\bf 
David Kastor\footnote{kastor@physics.umass.edu} and \c{C}etin  \c{S}ent\"{u}rk\footnote{csenturk@physics.umass.edu}} 

\vskip 0.5cm
{{Amherst Center for Fundamental Interactions\\
Department of Physics, University of Massachusetts, Amherst, MA 01003
     }}
\vspace{6pt}
\end{center}
\vskip 0.2in
\par
\begin{center}
{\bf Abstract}
 \end{center}
\begin{quote}
Higher curvature Lovelock  gravity theories can have  a number of maximally symmetric vacua with different values of the curvature.  
Critical surfaces in the space of Lovelock couplings separate regions with different numbers of such vacua, and there exist symmetry breaking regions with no maximally symmetric vacua.  
Especially in such regimes, it is interesting to ask what reduced symmetry vacua may exist.  
We study this question, focusing on vacua that are products of maximally symmetric spaces. 
For low order Lovelock theories, we assemble a map of such vacua over the Lovelock coupling space, displaying different possibilities for vacuum symmetry breaking.  We see indications of interesting structure, with {\it e.g.} product vacua in Gauss-Bonnet gravity covering the entirety of the symmetry breaking regime in $5$-dimensions, but only a limited  portion of it in $6$-dimensions.

\vfill
\vskip 2.mm
\end{quote}
\hfill
\end{titlepage}

\newpage

\section{Introduction}\label{Sec:Intro}

Higher curvature gravitational interactions have been investigated in a great many physical contexts.  Among such models, the special class of Lovelock gravity theories \cite{Lovelock:1971yv} is distinguished via having field equations that depend only on the Riemann tensor, and not on its derivatives, and hence include at most second derivatives of the metric tensor.  This has a number of important consequences at both the classical and quantum levels.  For example, it leads to a Hamiltonian formulation in terms of the standard canonical gravitational degrees of freedom \cite{Teitelboim:1987zz} and to the absence of the ghost degrees of freedom that are typical of higher curvature theories \cite{Zwiebach:1985uq,Zumino:1985dp}.

Lovelock theories include a single interaction term at each higher curvature order, so that the Lagrangian in $n$ spacetime dimensions is given by $\mc{L}=\sum_{k=0}^{p} c_k\mc{L}_k$ with\footnote{The antisymmetrized Kronecker symbol used here has overall strength $k!$ and is defined by 
\be\nonumber
\delta^{\a_1\dots \a_k}_{\b_1\dots \b_k} = k!\delta^{[\a_1}_{\b_1}\dots\delta^{\a_k]}_{\b_k}
= k!\delta^{\a_1}_{[\b_1}\dots\delta^{\a_k}_{\b_k]}.
\ee}
\be\label{Love}
\mc{L}_k=\frac{1}{2^k}\,\d^{\a_1\b_1\dots\a_k\b_k}_{\mu_1\nu_1\dots\mu_k\nu_k}
R_{\a_1\b_1}^{\m_1\n_1}\dots R_{\a_k\b_k}^{\m_k\n_k},
\ee
with the upper limit of the sum given by  $p\equiv[(n-1)/2]$.
The term $\mc{L}_0$ gives the cosmological constant term in the action, while $\mc{L}_1$ gives the Einstein-Hilbert term and the $\mc{L}_k$ with $k\ge 2$ are higher curvature terms.
The Lagrangian truncates  because the interactions $\mc{L}_k$ vanish identically for $k>n/2$, while for $n$ even the variation of $\mc{L}_{n/2}$ gives a total divergence and hence does not contribute to the equations of motion\footnote{In spacetime dimension $n=2k$, the term $\mc{L}_k$ is the Euler density, whose integral over a compact manifold without boundary is topologically invariant}.  This truncation distinguishes between even and odd dimensions.  Moving up from an even dimension to the next higher odd dimension, a new Lovelock interaction is introduced.  However, no new term is introduced in even dimensions.
The coefficients $c_k$ are the couplings of the theory, and we will be interested in how the space of possible vacua of Lovelock gravity varies as a function of these couplings.

The simplest vacua of Lovelock theories are maximally symmetric ones, and  depending on the values of the couplings $c_k$,  Lovelock gravity in $n$-dimensions may have up to $p$ such vacua with distinct curvatures \cite{Boulware:1985wk,Wheeler:1985nh,Wheeler:1985qd}. 
Assuming that $n>1$,  the curvature of a
maximally symmetric spacetime has the form
\be\label{Riem}
R^{\mu\nu}_{\a\b}=K\d^{\mu\nu}_{\a\b},
\ee
where the constant $K$ is related to the scalar curvature according to $K= {1\over n(n-1)}R$.  The maximally symmetric spacetime is then either Minkowski (M), de Sitter (dS), or anti-de Sitter (AdS) spacetimes\footnote{If instead we considered Euclidean metrics then we would have respectively Euclidean ($\mathds{E}$), spherical (S), or hyperbolic (H) spaces corresponding to these ranges of the curvature constant.}, depending on whether  $K=0$, $K>0$, or $K<0$. 
When the Riemann tensor has the constant curvature form (\ref{Riem}), the Lovelock equations of motion  reduce to  a $p$th-order polynomial equation for $K$.  Only real roots of this equation correspond to physical vacua.   Therefore, while there exists  a region of coupling space with $p$ distinct maximally symmetric vacua,  there are also regions with fewer such vacua.  For $p$ odd there will  be at least one maximally symmetric vacuum.  However, for $p$ even there is a range of couplings such that no maximally symmetric vacuum exist.  We can think of the surfaces in coupling space that divide such regions, with different numbers of maximally symmetric vacua, as critical surfaces of the theory.

Consider, for example, Gauss-Bonnet gravity in dimensions $n\ge 5$ where only the couplings $c_k$ with $k=0,1,2$ are taken to be nonzero\footnote{This includes the full set of allowable couplings for dimensions $n=5,6$, while for $n>6$ the higher order couplings $c_k$ with $k=3,\dots,p$ are taken to be zero in Gauss-Bonnet gravity.}.  In this case, the equations of motion yield a quadratic equation for the curvature constant $K$.  For fixed values of $c_0$ and $c_1$, one finds that there is always either a maximum or minimum value of $c_2$, depending on the sign of $c_0$, beyond which maximally symmetric vacua no longer exist.   It is then natural to ask what vacua exist beyond this critical value of the coupling $c_2$.  Such vacua will necessarily have less than the maximal symmetry allowed in a given spacetime dimension.

In this paper, we will investigate a simple class of alternative vacua with reduced amounts of symmetry and examine how the number and existence of such vacua vary over the space of Lovelock couplings.  The vacua we consider are products of maximally symmetric space(times).  In $n$ spacetime dimensions, we will write this product in the form 
$\mc{K}_1^d\times\mc{K}_2^{n-d}$.   Here, the first factor $\mc{K}_1^d$ is a Lorentzian maximally symmetric spacetime of dimension $d$ with curvature constant $K_1$, while the second factor $\mc{K}_2^{n-d}$ is an $(n-d)$-dimensional maximally symmetric Euclidean space with curvature constant $K_2$.  The existence of such reduced symmetry vacua introduces a new set of regions and critical surfaces in coupling space.  We will see how these regions either extend, or fail to extend, the region in which maximally symmetric vacua exist.

Some special cases of such product vacua have  already been discussed in literature. The product vacua  $M^4\times S^{3}$  and more generally  $\mc{K}_1^4\times S^{3}$  in third order Lovelock gravity were studied in \cite{MuellerHoissen:1985mm}  and  \cite{Canfora:2008iu} respectively. The Nariai and (anti)-Nariai \cite{Nar,Dadhich:2000zg} type factorizations $dS_2\times S^{n-2}$ and $AdS_2\times H^{n-2}$ were investigated in Einstein gravity \cite{Cardoso:2004uz}  and  in Gauss-Bonnet gravity \cite{LorenzPetzold:1987ih}, while the Bertotti-Robinson \cite{Bertotti:1959pf,Robinson:1959ev} type factorization AdS$_2\times S^{n-2}$ was investigated  in Einstein gravity \cite{Cardoso:2004uz}, in generic Lovelock gravity \cite{Maeda:2011ii}, in pure Lovelock gravity theories \cite{Dadhich:2012zq}, and in $5$-dimensional quadratic gravity \cite{Clement:2013twa}.   The product $\mc{K}_1^2\times\mc{K}_2^{3}$ in $5$-dimensional Gauss-Bonnet gravity was considered by \cite{Canfora:2007ux,Izaurieta:2012fi}.
Related work has also appeared in references \cite{Canfora:2013xsa,Canfora:2014iga} which study dynamical compactification, in the broken symmetry regime of Gauss-Bonnet gravity, that are products of a $4D$ FRW spacetime with a compact space having a time dependent scale factor.   The dynamics of possible compactifications has also been studied in \cite{Chirkov:2014nua,Chirkov:2015kja,Pavluchenko:2015daa}.

This paper is organized as follows.  In Section (\ref{Sec:3rdLove}) we give some more details of Lovelock gravity.  In order to keep our analysis of product vacua tractable
we will restrict our analysis to at most cubic order interactions in the curvature.  In Section (\ref{Einsteinsection}), in order to orient the subsequent discussion, we recall the maximally symmetric and product vacua of Einstein gravity.  In Section (\ref{GBsection}) we look at the maximally symmetric and product vacua of Gauss-Bonnet gravity in $n=5$ and $n=6$ dimensions.  In Section (\ref{3rdordersection}), we study product vacua in third order Lovelock theory, making a further restriction of the couplings to keep the problem tractable.  Finally, we offer some concluding remarks in Section (\ref{conclude}).

\section{Low order Lovelock theory}\label{Sec:3rdLove}

In practice, in order to keep our analysis of product vacua tractable, we will restrict our attention to Lovelock theories including only the first few, relatively low order interaction terms.  Accordingly, we will work with the theory described by the action 
\be\label{3rdLove}
I=\frac{1}{16\pi G_n}\int d^{n}x\sqrt{-g}\l(-2\La_0+R+\a_2\mc{L}_2+\a _3\mc{L}_3\r),
\ee
where we have written the cosmological and Einstein-Hilbert terms in the action in their conventional forms, while leaving the $2$nd and $3$rd order Lovelock terms in the compact form (\ref{Love}).   The explicit form of the second order Gauss-Bonnet term, which is dynamically relevant in dimensions $n\ge 5$, is given by
\be
\mc{L}_2=R^2-4R_\a^\m R_\m^\a+R_{\a\b}^{\m\n}R_{\m\n}^{\a\b}.
\ee
The explicit form for the third order term,  which is relevant in dimensions $n\ge 7$, is unwieldy.  
The equations of motion of a general Lovelock theory are given by 
$\sum_{k=0}^{p} c_k\mc{G}^{(k)}{}^\m_{\n}=0$, where
\be
\mc{G}^{(k)}{}^\m_\n = -{1\over 2^{k+1}}\,
\d^{\m\a_1\b_1\dots\a_k\b_k}_{\n\s_1\k_1\dots\s_k\k_k}R_{\a_1\b_1}^{\s_1\k_1}\dots R_{\a_k\b_k}^{\s_k\k_k}.
\ee
and the expression for ${\mc{G}^{(1)}}^\mu_\n$ reproduces the ordinary Einstein tensor $G^\mu_\n$.
For our theory (\ref{3rdLove}) the equations of motion are then given by
\be\label{3rdLoveEqn}
\La_0\d^\mu_\nu+G^\mu_\nu+\a_2{\mc{G}^{(2)}}^\mu_\nu+\a_3{\mc{G}^{(3)}}^\mu_\nu=0.
\ee
In order to orient the discussion below, we will first set the couplings $\alpha_2=\alpha_3=0$ and look at product vacua of Einstein gravity.  We will then take $\alpha_2$ to be nonvanishing and study the problem for Gauss-Bonnet gravity\footnote{In the context of the low energy effective action from string theory, it is noted in \cite{Boulware:1985wk} that $\alpha_2>0$, but we will consider all possible values here.}.  Finally, we will allow $\alpha_3$ to be nonzero, although we will restrict our attention to a subclass of theories with a definite relation between the couplings $\alpha_2$ and $\alpha_3$, in order to make the analysis manageable.

\section{Product vacua in Einstein gravity}\label{Einsteinsection}

\begin{figure}[t]
\centering
\includegraphics[width=1.6in,angle=270]{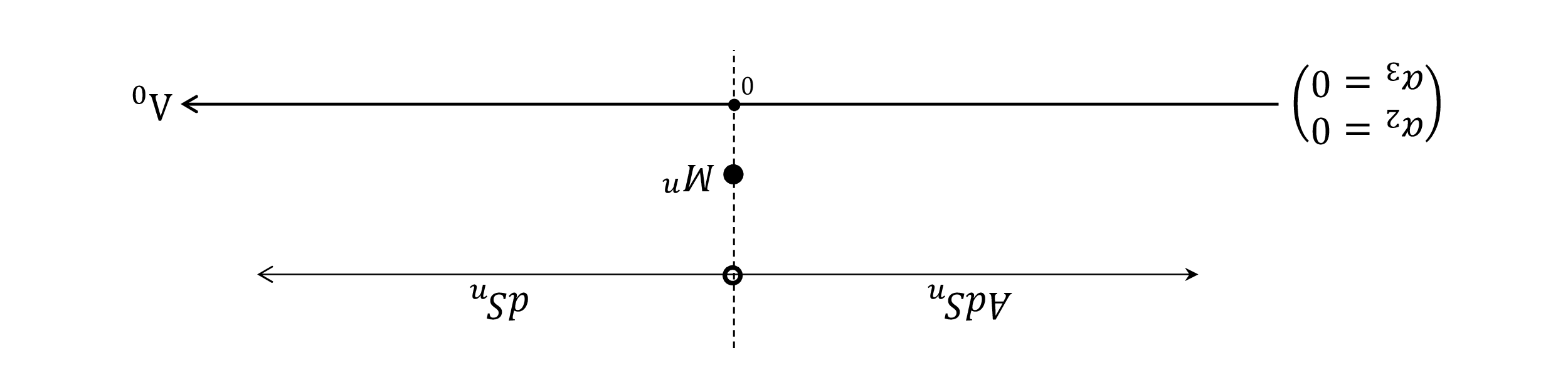}
\caption{\sl Maximally symmetric vacua of Einstein gravity in $n$ dimensions.}
\label{Fig1}
\end{figure}

In order to orient the discussion of product vacua in Lovelock theories, we first present the analysis for Einstein gravity, setting the $\alpha_2=\alpha_3=0$ in our theory (\ref{3rdLove}).  Our theory is then parameterized by the cosmological constant $\Lambda_0$.  For each value of the cosmological constant, the theory has a maximally symmetric vacuum, with curvature constant $K$ related to the cosmological constant by
\be\label{constant}
K=\frac{2\Lambda_0}{(n-1)(n-2)}.
\ee
The different possible ranges for the cosmological constant $\Lambda_0<0$, $\Lambda_0=0$ and $\Lambda_0>0$ correspond to AdS, Minkowski and dS vacua respecitvely.
We illustrate this situation in Figure (\ref{Fig1}), which serves as the prototype for subsequent, more complicated figures which we will use to display our results.  In this case, the figure shows the different types of maximally symmetric vacua corresponding to different values of the cosmological constant.  There is no evidence of critical behavior in this case.  As noted above, for Einstein gravity there is a unique maximally symmetric vacuum associated with each value of the cosmological constant.

We now consider certain reduced symmetry vacua of Einstein gravity, taking
the $n$-dimensional spacetime manifold to be a direct product of two maximally symmetric submanifolds $\mc{K}_1^d$ having and $\mc{K}_2^{n-d}$ of dimensions $d$ and $n-d$ respectively.  We will assume for the moment that $n\ge 4$ and that $n-1> d >1$ and consider the case of $1$-dimensional submanifolds separately.
The metric is then taken to have the form
\be
ds^2=g_{\m\n}(x)dx^\m dx^\n=g_{ab}(u)du^adu^b+g_{ij}(v)dv^idv^j,
\ee
where $g_{ab}(u)$ is the metric on $\mc{K}_1^d$, which is assumed to have Lorentzian signature, and $g_{ij}(v)$ is the metric on the manifold $\mc{K}_2^{n-d}$, which is taken to have Euclidean signature. Coordinate indices along $\mc{K}_1^d$ have been denoted by $a,b,c,\ldots$ and coordinates indices for $\mc{K}_2^{n-d}$ by  $i,j,k,\ldots$. This construction enables us to decompose the Riemann tensor $R^{\m\n}_{\a\b}$ of the full spacetime as
%
\begin{figure}[t]
\centering
\includegraphics[width=1.6in,angle=270]{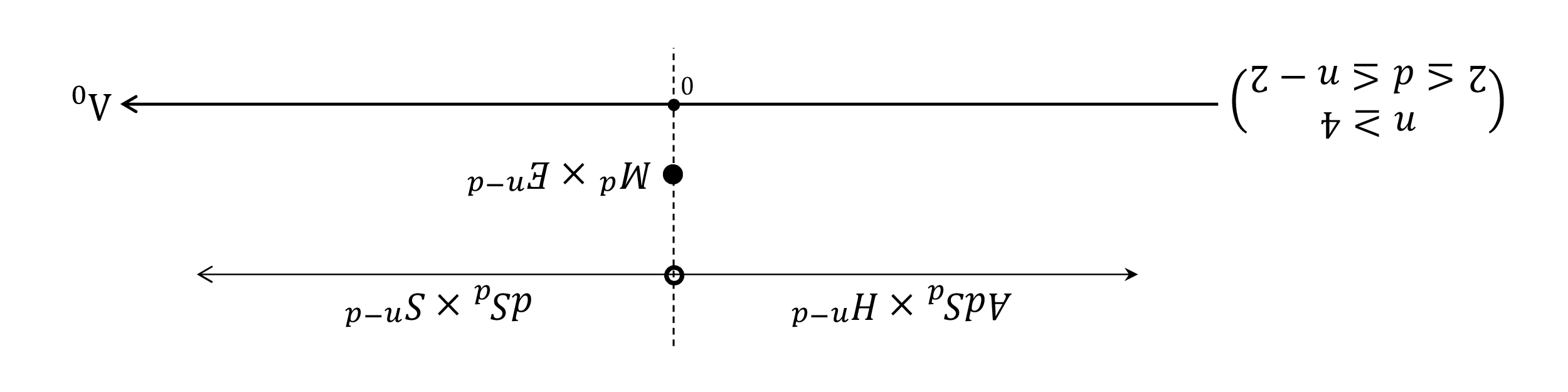}
\caption{\sl Direct product vacua of Einstein gravity in $n$ dimensions.}
\label{Fig5}
\end{figure}
\be
R^{ab}_{cd}=R^{\quad  ab}_{(1)cd},~~R^{ij}_{kl}={R}{}^{\quad ij}_{(2)kl},~~R^{ai}_{bj}=0,
\ee
where ${R}{}^{\quad ab}_{(1)cd}$ and ${R}{}^{\quad ij}_{(2)kl}$ are the Riemann tensors of $\mc{K}_1^d$ and $\mc{K}_2^{n-d}$, respectively, which are each assumed to have the constant curvature form
\be\label{dRiem}
R^{\quad ab}_{(1)cd}=K_1\d^{ab}_{cd},\qquad 
R^{\quad ij}_{(2)kl}=K_2\d^{ij}_{kl}
\ee
with curvature constants $K_1$ and $K_2$.  Plugging into the Einstein equation then yields for these constants
\be\label{prodcurv}
K_1 = {2\Lambda_0\over (d-1)(n-2)},\qquad K_2 = {2\Lambda_0\over (n-d-1)(n-2)}
\ee
We see in particular that $K_1$ and $K_2$ necessarily each have the same sign as the cosmological constant $\Lambda_0$.   The product vacua we consider, therefore, are always of the form $AdS_d\times H^{n-d}$ for $\Lambda_0$ negative, $dS_d\times S^{n-d}$ for $\Lambda_0$ positive, and $M_d\times E^{n-d}$ for vanishing cosmological constant.  This result is illustrated in Figure (\ref{Fig5}).  As with the maximally symmetric vacua, there is no evidence of critical behavior for the product vacua.  Such vacua exist for all values of the cosmological constant.

\begin{figure}[t]
\centering
\includegraphics[width=1.6in,angle=270]{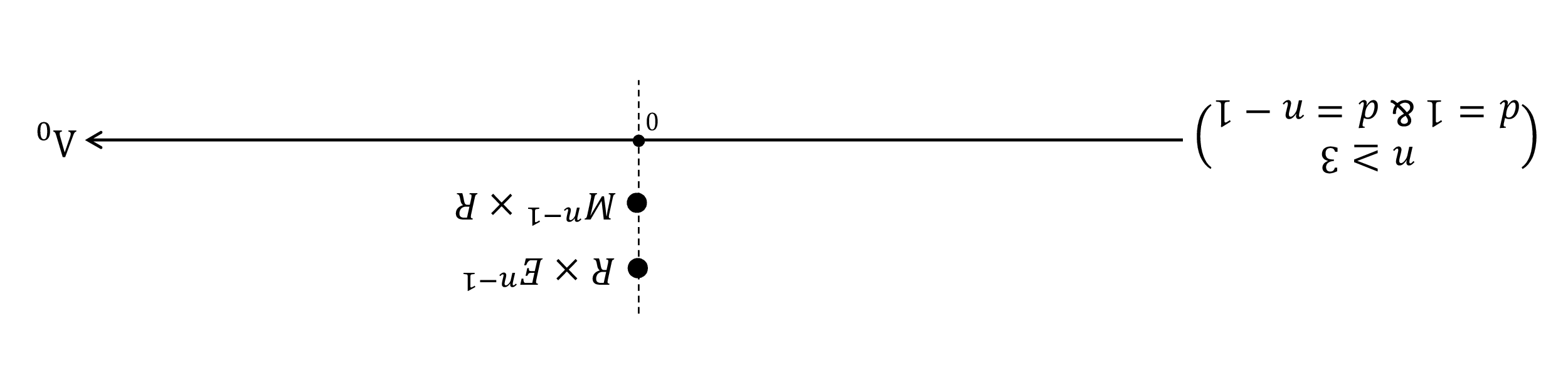}
\caption{\sl Direct product vacua of Einstein gravity in the case of one-dimensional submanifolds in $n$ dimensions, which exist only for $\Lambda_0=0$.}
\label{Fig4}
\end{figure}

The cases $d=1$ and $d=n-1$, where one of the submanifolds $\mc{K}_1^d$ or  $\mc{K}_2^{n-d}$ has dimension $1$ and is therefore flat,  require special handling.  In these cases we see that one or the other of the terms in the formal solution in (\ref{prodcurv}) diverges.  This can be traced back to an inconsistency between the different components of the Einstein equations, which implies that there is no product vacuum unless $\Lambda_0=0$, as illustrated  in Figure (\ref{Fig4}).

\section{Product vacua in Gauss-Bonnet gravity}\label{GBsection}

Now equipped with an understanding of product vacua in Einstein gravity, we move on to consider such vacua in Gauss-Bonnet gravity, which is obtained by setting the coupling 
$\alpha_3=0$ in the action (\ref{3rdLove}).  

\subsection{Maximally symmetric vacua}
Let us first consider the maximally symmetric vacua, having constant curvature (\ref{Riem}).  We will characterize these solutions by an effective cosmological constant $\Lambda$ related to the curvature constant $K$ by
\be
\Lambda = {(n-1)(n-2)\over 2} K,
\ee
which is the relation that holds between the curvature and cosmological constants for constant curvature solutions in Einstein gravity (\ref{constant}).  It is well known that the equations of motion for constant curvature solutions in Gauss-Bonnet gravity reduce to a quadratic equation, which is given in terms of the effective cosmological constant by
\be\label{VacEqnGB}
\tilde{\a}_2\La^2+\La-\La_0=0,
\ee
where $\tilde\alpha_2= {2(n-3)(n-4)\over (n-1)(n-2)}\alpha_2$.  The maximally symmetric solutions are then characterized by the effective cosmological constants
\be\label{VacGB}
\La_{\pm}=-\frac{1}{2\tilde{\a}_2}\l(1\pm\sqrt{1+4\tilde{\a}_2\La_0}\r).
\ee
Since only real values of $\Lambda$ correspond to physical vacua, the number of such solutions will depend on the cosmological constant $\Lambda_0$ and the coupling strength 
$\alpha_2$ of the Gauss-Bonnet term.  For sufficiently small values of $\tilde\alpha_2$ there will always be two physical maximally symmetric vacua.  In the limit of small Gauss-Bonnet coupling, such that $|4\tilde{\a}_2\La_0|\ll 1$, these are given approximately by
\be
\Lambda_- \simeq \Lambda_0,\qquad \Lambda_+\simeq -{1\over \tilde\alpha_2}
\ee
One sees that $\Lambda_-$ matches on to the vacuum of Einstein gravity in this limit, while the $\Lambda_+$ branch goes off to infinite curvature.  For this reason, the corresponding branches of solutions in (\ref{VacGB}) are known as the Einstein and Gauss-Bonnet branches respectively.

\begin{figure}[t]
\centering
\includegraphics[width=4.35in,angle=270]{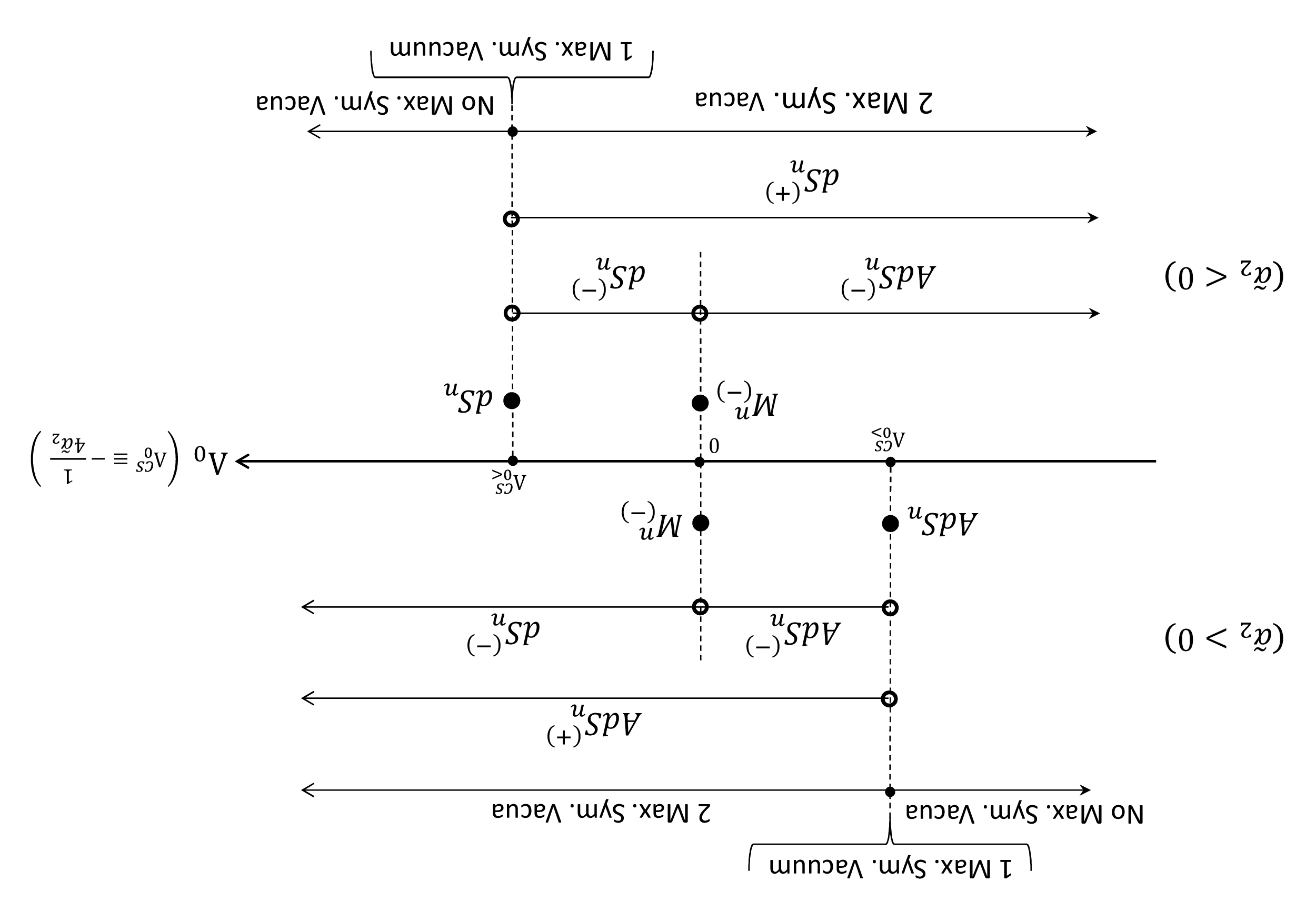}
\caption{\sl Maximally symmetric vacua of Gauss-Bonnet gravity in $n$ dimensions. Here, the plus and minus signs refer to the branches in the solution (\ref{VacGB}), and $\La_{0>}^{CS}$ and $\La_{0<}^{CS}$ represent the CS points when $\tilde{\a}_2>0$ and $\tilde{\a}_2<0$, respectively.}
\label{Fig2}
\end{figure}

The set of maximally symmetric vacua of Gauss-Bonnet gravity is displayed in Figure (\ref{Fig2}).  To read the figure,  envision fixing a value of the Gauss-Bonnet coupling 
$\alpha_2$, or equivalently $\tilde\alpha_2$, and asking how the number and character of the maximally symmetric vacua vary with the value of the cosmological constant $\Lambda_0$.
The behavior is qualitatively the same for all values of $\tilde\alpha_2>0$, which are displayed in the top half of the diagram, and also for all values of $\tilde\alpha_2<0$, which are displayed on the bottom half.  For $\tilde\alpha_2>0$, it follows from (\ref{VacGB}) that two distinct vacua exist for all values of $\Lambda_0>\La_{0}^{CS}$, while at the critical value 
\be\label{CS}
\La_{0}^{CS} = -{1\over 4\tilde\alpha_2},
\ee
the two branches of solutions become degenerate and a single unique maximally symmetric vacuum exists.  We call this critical point the CS point because in $n=5$ dimensions the theory can be re-expressed as a Chern-Simons theory when the Gauss-Bonnet coupling and cosmological constant are related in this way (see reference \cite{Crisostomo:2000bb}).
It also follows from (\ref{VacGB}) that no physical vacua exist for $\Lambda_0<\La_{0}^{CS}$.  This implies that whatever vacuum solutions exist in this region of coupling space will necessarily be symmetry breaking ones.

We can also examine the character of the maximally symmetric vacua on the two branches for $\Lambda_0>\La_{0}^{CS}$.  One finds that the sign of the effective cosmological constant 
$\Lambda_-$ for the Einstein branch is precisely correlated with the cosmological constant $\Lambda_0$, so that the solutions along this branch are dS, Minkowski, or AdS depending on whether the value of $\Lambda_0$ is positive, zero, or negative.  These are respectively the $dS_n^{(-)}$, $M^n_{(-)}$ and $AdS_n^{(-)}$ solutions shown in the top half of Figure (\ref{Fig2}).  
Note that $\La_{0}^{CS}\rightarrow -\infty$ in the limit of vanishing Gauss-Bonnet coupling, and that the Einstein branch correctly reduces to the result in Figure (\ref{Fig1}).
On the Gauss-Bonnet branch, however, the effective cosmological constant is always negative, independent of the sign of 
$\Lambda_0$, giving the $AdS_n^{(+)}$ branch of vacua.  Finally, this entire structure is mirrored in a straightforward way for $\tilde\alpha_2<0$ on the bottom of Figure (\ref{Fig2}).

\subsection{Product vacua}

We now turn to vacua which, as above, are products $\mc{K}_1^d\times \mc{K}_2^{n-d}$ of maximally symmetric submanifolds.   The Gauss-Bonnet equations of motion for such product vacua reduce to a coupled set of quadratic equations for the curvature constants $K_1$ and $K_2$ given by
\begin{align}\label{gb1}
\alpha_2\left( {(d-1)!\over (d-5)!}K_1^2  + {2(d-1)!D!\over (d-3)!(D-2)!} K_1K_2\right.&\left. + {D!\over (D-4)!}K_2^2\right)\\ & +{(d-1)!\over (d-3)!}K_1+ {D!\over (D-2)!}K_2-2\Lambda_0 =0\nonumber
\end{align}
\begin{align}\label{gb2}
\alpha_2\left( {(D-1)!\over (D-5)!}K_2^2 + {2(D-1)!d!\over (D-3)!(d-2)!}K_1K_2\right.&\left.+ {d!\over (d-4)!}K_1^2\right)\\ &+{(D-1)!\over (D-3)!}K_2+ {d!\over (d-2)!}K_1-2\Lambda_0 =0\nonumber
\end{align}
where $D=n-d$.   The linear equations that result from setting the Gauss-Bonnet coupling $\alpha_2$ to zero were used above to obtain the product solutions in Einstein gravity (\ref{prodcurv}).  However, for $\alpha_2\neq 0$ we cannot write down a general analytic solution to the equations.  For sufficiently small values of $d$ or $D$, however, the equations simplify and yield interesting results, and we will focus on such cases.  In particular, we will focus on product vacua for Gauss-Bonnet gravity in $n=5$ dimensions, which is the lowest dimension in which the Gauss-Bonnet term is relevant, and in $n=6$ dimensions, where it is also the highest order Lovelock term.  The subsequent term $\mc{L}_3$ which is cubic in the curvature becomes relevant in $n=7$ dimensions.

In $n=5$ dimensions we consider the cases of $3+2$ and $2+3$ dimensional splits, which differ only in which factor of the product $\mc{K}_1^d\times \mc{K}_2^{n-d}$ is Lorentzian and which is Euclidean.  Taking the case $d=3$, $D=2$ case first, the equations of motion simplify to
\begin{align}\label{GBeqns}
8\alpha_2 K_1K_2 +2(K_1+K_2)-2\Lambda_0&=0\\
6K_1-2\Lambda_0&=0\nonumber
\end{align}
The resulting curvature constants can then be written in the form
\be\label{GBprods}
K_1 = {\Lambda_0\over 3},\qquad 
K_2 = {1\over \left(1-{\Lambda_0\over\,\,\,\Lambda_0^{CS}}\right)}\cdot{2\over 3}\Lambda_0,
\ee
where $\Lambda_0^{CS}=-{3\over 4\alpha_2}$ is the CS value of the cosmological constant (\ref{CS}) in $n=5$ dimensions.
These values of the curvature constants $K_1$ and $K_2$ approach those for Einstein gravity (\ref{prodcurv}) in the limit of small Gauss-Bonnet coupling.

\begin{figure}[t]
\centering
\includegraphics[width=4.0in,angle=270]{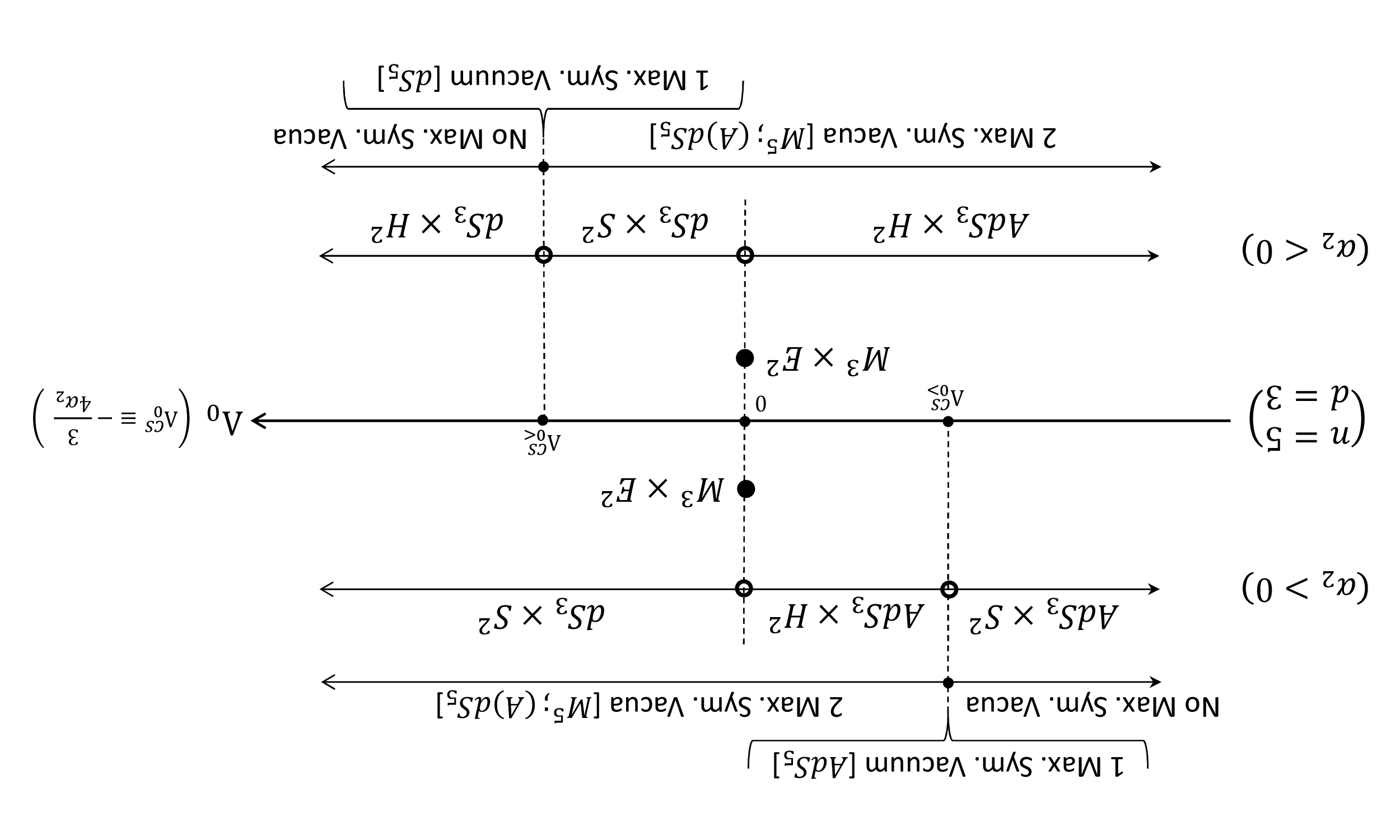}
\caption{\sl Direct product vacua of Gauss-Bonnet gravity when $n=5$ and $d=3$.}
\label{Fig8}
\end{figure}

We display these results in Figure (\ref{Fig8}).  Again, to read this diagram, envision fixing a non-zero value of the Gauss-Bonnet coupling $\alpha_2$ and then consider the full range of values for the cosmological constant $\Lambda_0$.  Let us focus on the top half of the diagram where $\alpha_2>0$.  
As in the case of Einstein gravity, displayed above in Figure (\ref{Fig5}), there is again a single product vacuum across the full range of values for the cosmological constant.  However, the curvatures of the two factor spaces are no longer precisely correlated with the sign of $\Lambda_0$ across the full range of values, as they are in the Einstein case.   For $\Lambda_0>\Lambda_0^{CS}$ there is such a precise correlation, with both $K_1$ and $K_2$ being either positive, zero or negative in correspondence with the value of 
$\Lambda_0$.  This produces, in succession for smaller values of $\Lambda_0$, products solutions of the form $dS_3\times S^2$, $M^3\times E^2$ and $AdS_3\times H^2$.
However, the solutions (\ref{GBprods}) have a critical point at $\Lambda=\Lambda_0^{CS}$, where the curvature $K_2$ diverges.  For $\Lambda<\Lambda_0^{CS}$ the curvature of the Euclidean space $\mc{K}_2^{2}$ becomes positive, opposite to the sign of $\Lambda_0$, giving a product of the form $AdS_3\times S^2$.  This behavior is mirrored for $\alpha_2<0$ on the bottom half of the diagram.

\begin{figure}[t]
\centering
\includegraphics[width=3.75in,angle=270]{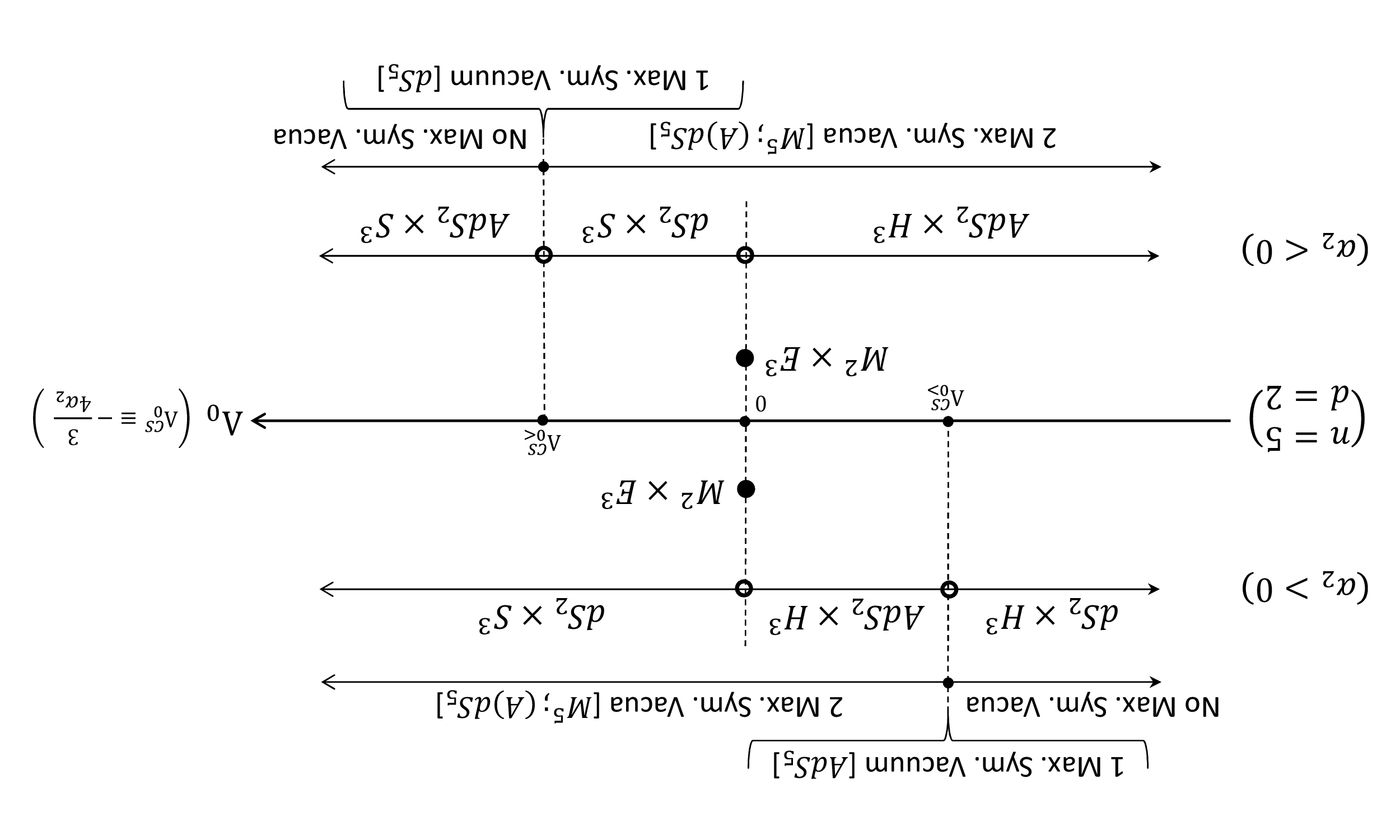}
\caption{\sl Direct product vacua of Gauss-Bonnet gravity when $n=5$ and $d=2$.}
\label{Fig7}
\end{figure}

The case in $n=5$ dimensions for a product of a maximally symmetric $d=2$ dimensional Lorentzian spacetime with a maximally symmetric $D=3$ dimensional Euclidean space is very similar.  The equations of motion (\ref{gb1}) and (\ref{gb2}) again reduce to have the drastically simplified (\ref{GBeqns}), but with the curvature constants swapped, so that the solutions are now given by
\be\label{GBprods2}
K_1 = {1\over \left(1-{\Lambda_0\over\,\,\,\,\,\Lambda_0^{CS}}\right)}\cdot{2\over 3}\Lambda_0,\qquad 
K_2     = {\Lambda_0\over 3}
\ee
These results are displayed in Figure (\ref{Fig7}), which is very similar to Figure (\ref{Fig8}), the key difference being that moving through the CS value of the cosmological constant changes the sign of the curvature of the Lorentzian, rather than the Euclidean, part of the product in this case.

It is intriguing that the critical point occurs at $\Lambda_0=\Lambda_0^{CS}$ for both these product vacua, this being the same as the critical point that separates different regimes for maximally symmetric vacua, as displayed in Figure (\ref{Fig2}).  If via some mechanism, the cosmological constant were dynamical, then various possibilities for vacuum transitions exist.  Assume for concreteness that the Gauss-Bonnet coupling is positive.    
If $\Lambda_0$ then evolved through the CS point from above, a maximally symmetric $AdS_5$ vacuum of Figure (\ref{Fig2}) might transition into the $AdS_3\times S^2$ vacuum of Figure (\ref{Fig8}), a process of spontaneous compactification.

We now move on to discuss product vacua for Gauss-Bonnet gravity in $n=6$ spacetime dimensions, which we will see display different types of critical behavior.  We will consider $3+3$, $4+2$ and $2+4$ splits into Lorentzian and Euclidean factors in turn.  In $n=5$ dimensions we saw that the product vacua existed for all values of the cosmological constant, providing possible vacuum states in the broken symmetry regime.  We will see that this is no longer the case in $n=6$ dimensions.

\begin{figure}[t]
\centering
\includegraphics[width=4.0in,angle=270]{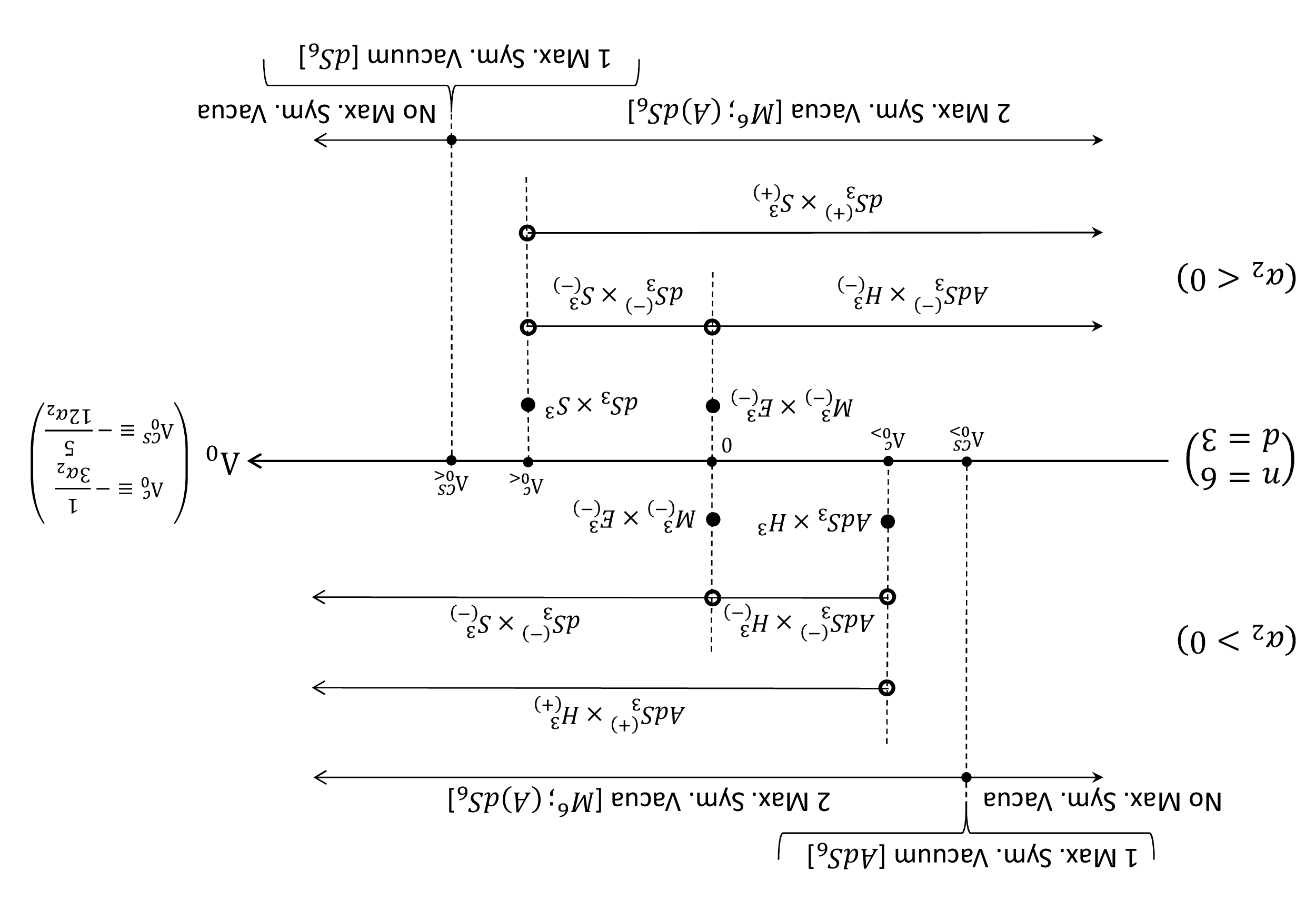}
\caption{\sl Direct product vacua of Gauss-Bonnet gravity when $n=6$ and $d=3$.}
\label{Fig10}
\end{figure}

For a $3+3$ split one finds that the equations of motion (\ref{gb1}) and (\ref{gb2}) simplify to
\begin{align}
24\alpha_2 K_1K_2 + 2K_1 +6K_2-2\Lambda_0 &=0\\
24\alpha_2 K_1K_2 + 2K_2 +6K_1-2\Lambda_0 &=0
\end{align}
which has the two solutions $K_1=K_2=K_\pm$ where 
\be
K_\pm = {\Lambda_0^c\over 2}\left(1\pm\sqrt{1-{\Lambda_0\over\Lambda_0^c}}\right)
\ee
and $\Lambda_0^c = -{1\over 3\alpha_2}$ is a new critical point that arises in this system.   Focusing on $\alpha_2>0$, which is illustrated in the top half of Figure (\ref{Fig10}), there will be two physical $3+3$ product solutions for $\Lambda_0>\Lambda_0^c$ and none in the regime $\Lambda_0<\Lambda_0^c$, with a unique solution at the critical point.  In $n=6$ dimensions, one finds from (\ref{CS}) that 
$\Lambda_0^{CS}=-{5\over 12\alpha_2}$, so that with $\alpha_2>0$ one has the ordering $\Lambda_0^{CS}<\Lambda_0^c$.  It follows that in the symmetry breaking regime $\Lambda<\Lambda_0^{CS}$, where no maximally symmetric vacua exist, there are also no $3+3$ split product vacua.  One also finds that, as for the maximally symmetric solutions, there is an ``Einstein" branch of product solutions with $K_1=K_2=K_-$ which approaches the analogous product vacua of Einstein gravity in the limit of small Gauss-Bonnet coupling, and a ``Gauss-Bonnet" branch with $K_1=K_2=K_+$ where the curvatures of both factors diverge in this limit.  On the Einstein branch, the curvatures of both factors are precisely correlated with the sign of $\Lambda_0$, while on the Gauss-Bonnet branch both factors are always negatively curved.  Finally, this whole structure is mirrored on the bottom half of the diagram for $\alpha_2<0$.

\begin{figure}[t]
\centering
\includegraphics[width=4.0in,angle=270]{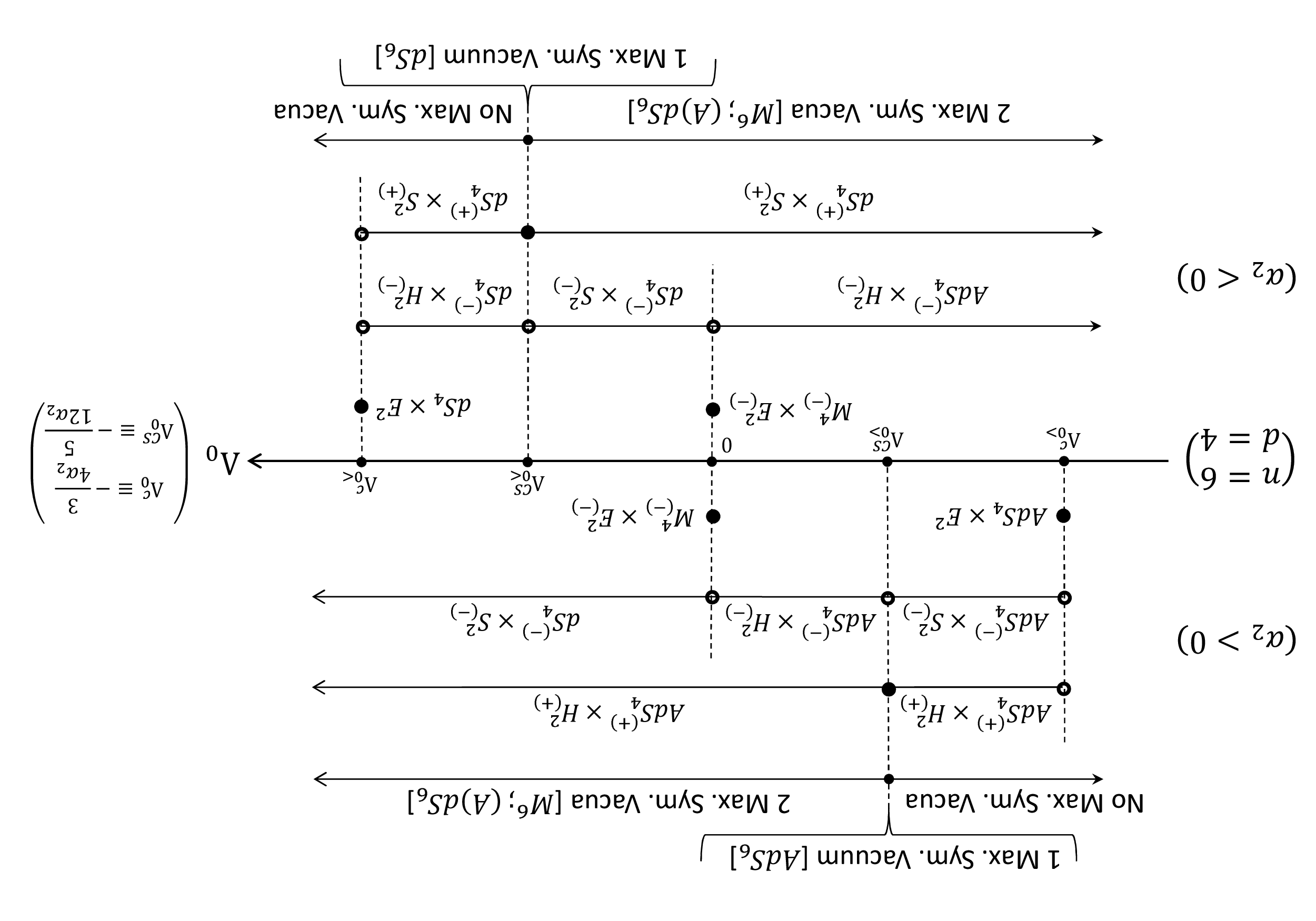}
\caption{\sl Direct product vacua of Gauss-Bonnet gravity when $n=6$ and $d=4$.}
\label{Fig11}
\end{figure}

\begin{figure}[t]
\centering
\includegraphics[width=4.0in,angle=270]{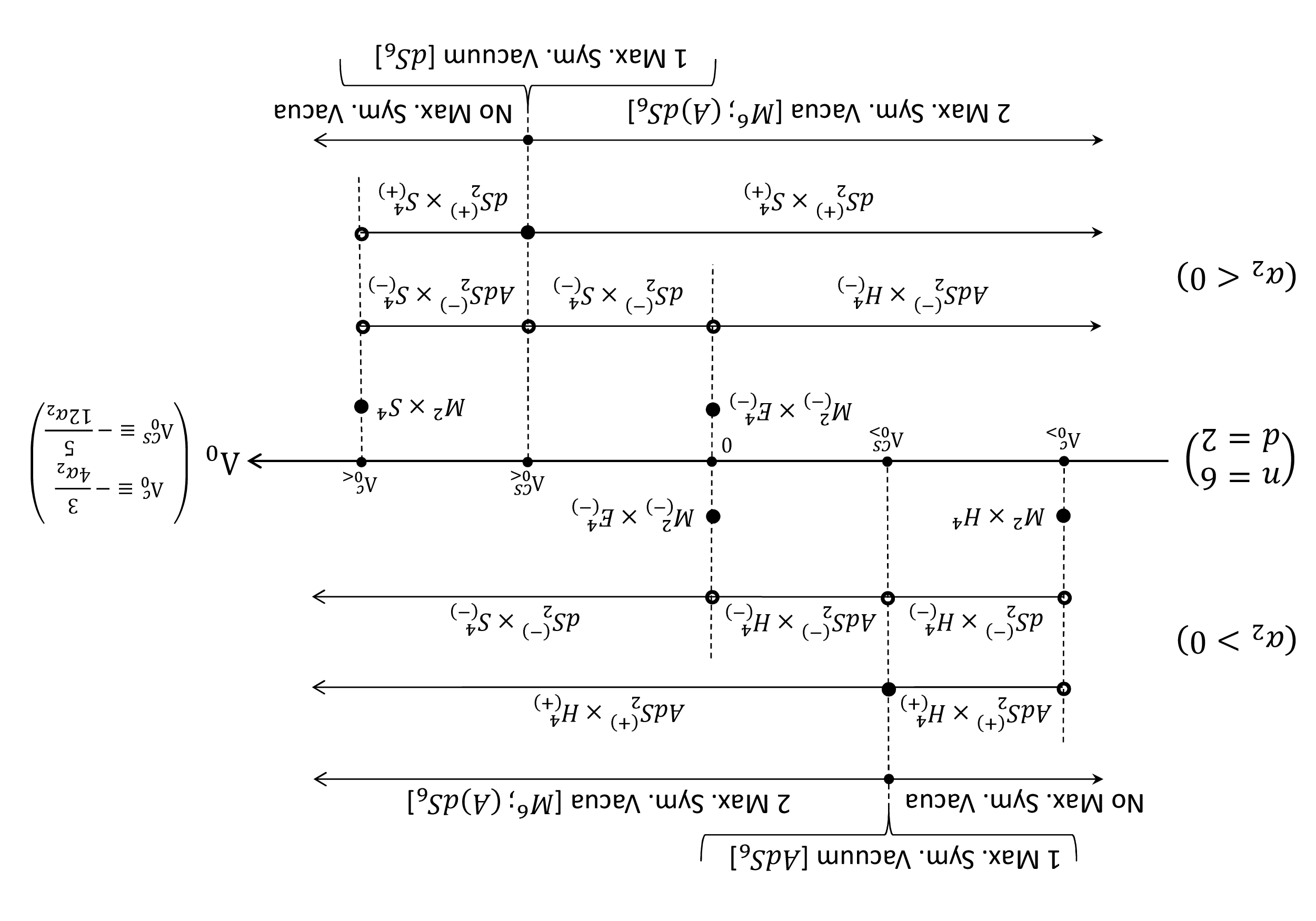}
\caption{\sl Direct product vacua of Gauss-Bonnet gravity when $n=6$ and $d=2$.}
\label{Fig9}
\end{figure}

The situation is similar in most respects for the $d=4$, $D=2$ product vacua.  In this case the equations (\ref{gb1}) and (\ref{gb2}) for the curvature constants $K_1$ and $K_2$ reduce to
\begin{align}\label{42split1}
24 \alpha_2 K_1K_2 +6K_1+2K_2-2\Lambda_0 &=0\\
24\alpha_2 K_1^2 +12 K_1-2\Lambda_0 &=0\label{42split2}
\end{align}
which have the solutions
\be\label{4plus2}
K_1^\pm = {\Lambda_0^c\over 3}B^\pm,\qquad K_2^\pm =\Lambda_0^c{B^\pm(B^\pm-1)\over 3B^\pm-1}
\ee
where in this case $\Lambda_0^c = -{3\over 4\alpha_2}$ and $B^\pm = \left(1\pm\sqrt{1-{\Lambda_0\over \Lambda_0^c}}\right) $.  These solutions are displayed in Figure (\ref{Fig11}).  Focusing on the top half of the diagram with $\alpha_2>0$, there are again two branches of solutions for all values of the cosmological constant 
$\Lambda_0>\Lambda_0^c$.  However, in this case we have $\Lambda_0^{CS}>\Lambda_0^c$, so that the product vacua extend some distance into the broken symmetry regime.
The curvatures $K_1^+$ and $K_2^+$ are always negative along the Gauss-Bonnet branch, so these solutions are $AdS_4\times H^2$.  The Einstein branch is more intricate.  
For $\Lambda_0>\Lambda_0^{CS}$, the signs of the curvatures $K_1^-$ and $K_2^-$ are both correlated with the sign of $\Lambda_0$ as they were in the $3+3$ split product described above.  However, the denominator of the expression for $K_2^-$ vanishes linearly\footnote{One finds that the factor in the denominator of $K_2^-$ in (\ref{4plus2}) can be written as
\be 3B^--1 = -5 \left({1-{\Lambda_0\over\,\,\,\,\, \Lambda_0^{CS}}\over 2+3\sqrt{1-{\Lambda_0\over \Lambda_0^c}}}\right).\ee}
 at $\Lambda_0=\Lambda_0^{CS}$ leading to a transition to $AdS_4\times S^2$ in the range $\Lambda_0^c<\Lambda_0<\Lambda_0^{CS}$.   Precisely at the critical point $\Lambda_0=\Lambda_0^c$ there is a single $AdS_4\times E^2$ solution.

The case of a  $2+4$ split product vacua in $n=6$ dimensions again reduces to equations (\ref{42split1}) and (\ref{42split2}), but now with the curvatures $K_1$ and $K_2$ swapped.  The resulting configurations are shown in Figure (\ref{Fig9}).  These vacua also exist a finite distance into the broken symmetry regime.

\begin{figure}[t]
\centering
\includegraphics[width=3.75in,angle=270]{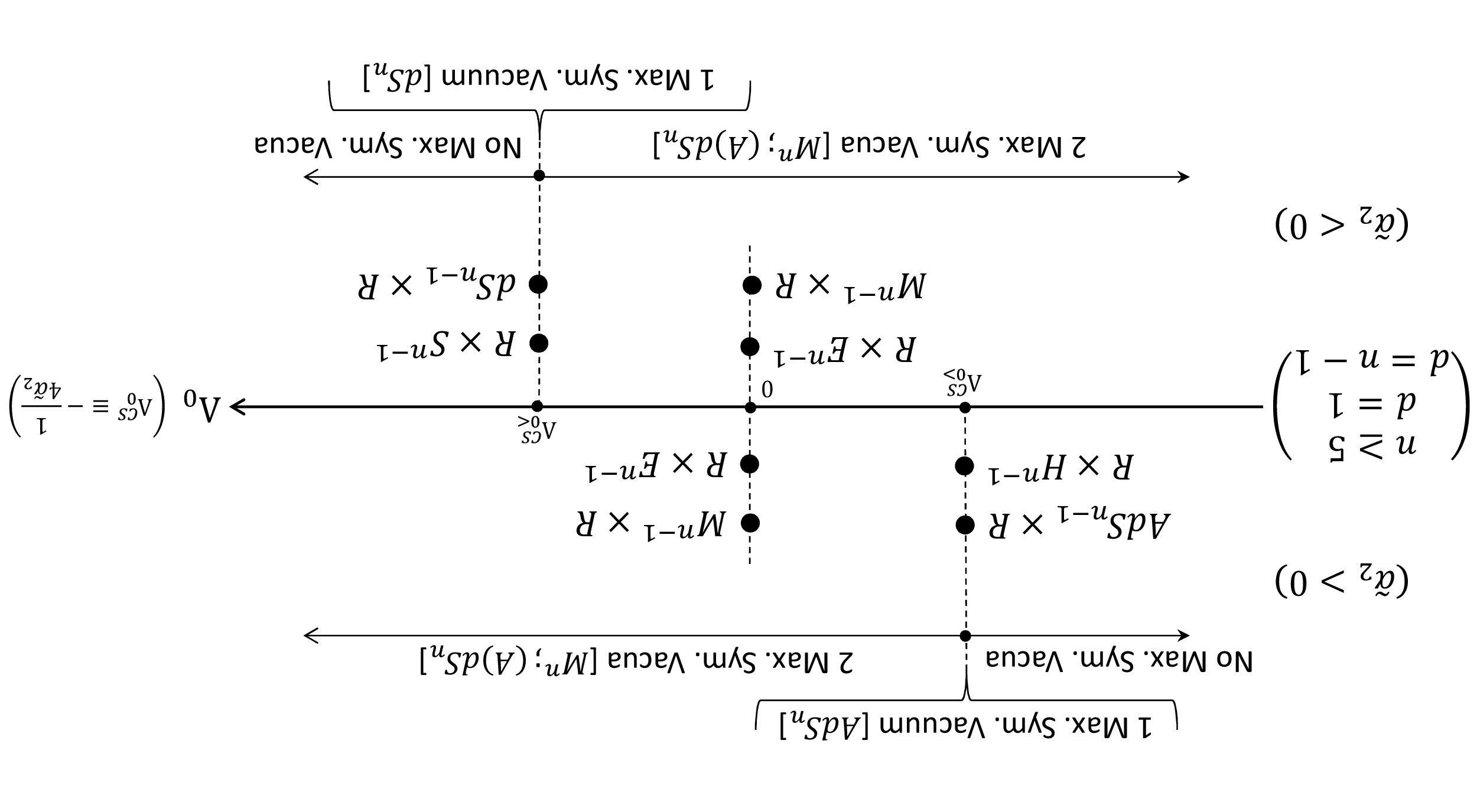}
\caption{\sl Direct product vacua of Gauss-Bonnet gravity in the case of one-dimensional submanifolds in $n$ dimensions.}
\label{Fig6}
\end{figure}

Finally, as in Einstein gravity, the cases  with $1$-dimensional factors, $d=1$ and $d=n-1$, require special handling, although in this case we are able to do the analysis for general spacetime dimension $n$.  Taking $d=1$, the equations of motion (\ref{gb1}) and (\ref{gb2}) reduce to 
\begin{align}
\alpha_2(n-1)(n-2)(n-3)(n-4)K_2^2 +(n-1)(n-2)K_2-2\Lambda_0 &=0  \\
\alpha_2(n-2)(n-3)(n-4)(n-5)K_2^2 +(n-2)(n-3)K_2-2\Lambda_0 &=0
\end{align}
These equations are inconsistent, except for the two special cases
\begin{align}
\Lambda_0 &=0,\qquad K_2=0\\
\Lambda_0 &=\Lambda_0^{CS},\qquad K_2={4\over (n-1)(n-2)}\Lambda_0^{CS}
\end{align}
which are displayed along with the corresponding results for $d=n-1$ in Figure (\ref{Fig6}).  This is a curious result.  Recall that in Einstein gravity, as shown in Figure (\ref{Fig4}), there are no similar $1$-dimensional product vacua with a non-zero value of the coupling constant.  However, any value of the cosmological constant in Einstein gravity can be thought of as being a CS value in the sense considered here.  It would be interesting to look at the
CS limits of higher order Lovelock theories to see if any pattern emerges with respect to product vacua with flat directions\footnote{Note that warped products with one dimensional factors were found for Lovelock theories with CS couplings in \cite{Kastor:2006vw}.}.

\section{Vacua in third order Lovelock gravity}\label{3rdordersection}

We now consider the full `low order' Lovelock theory introduced in Section (\ref{Sec:3rdLove}), which includes the third order Lovelock term as well.
This will be relevant in $n=7$ dimensions and beyond.  The equation determining the maximally symmetric vacua is now given by
\be\label{cubic}
\tilde{\a}_3\La^3+\tilde{\a}_2\La^2+\La-\La_0=0,
\ee
where $\tilde\alpha_2$ is as given above and $\tilde{\a}_3=\frac{4(n-3)(n-4)(n-5)(n-6)}{(n-1)^2(n-2)^2}\,\a_3$.
Since this is a cubic equation, there will always be at least one real root and, hence, at least one physical maximally symmetric vacuum state.
Therefore, third order Lovelock theory has no broken symmetry regime in coupling space.
The roots of  equation (\ref{cubic}) are given by
\bea
\La_{1}&=&\frac{1}{3\tilde{\a}_3}\l[-\tilde{\a}_2+A+\frac{\D_0}{A}\r],\label{Rroot}\\
\La_{2\pm}&=&-\frac{1}{6\tilde{\a}_3}\l[2\tilde{\a}_2+A+\frac{\D_0}{A}\pm i\sqrt{3}\l(A-\frac{\D_0}{A}\r)\r],\label{Croots}
\eea
where $A^3=\1o2\l[-\D_1+\sqrt{\D_1^2-4\D_0^3}\r]$
with $\D_0=\tilde{\a}_2^2-3\tilde{\a}_3$ and $\D_1=2\tilde{\a}_2^3-9\tilde{\a}_2\tilde{\a}_3-27\tilde{\a}_3^2\La_0$.
As noted in \cite{Amirabi:2013bqa}, when one takes the Gauss-Bonnet limit, $\alpha_3\rightarrow 0$, the real root $\La_{1}$ diverges, while the complex conjugate pair $\La_{2\pm}$ become the solutions (\ref{VacGB}).

\begin{figure}[t]
\centering
\includegraphics[width=3.0in,angle=270]{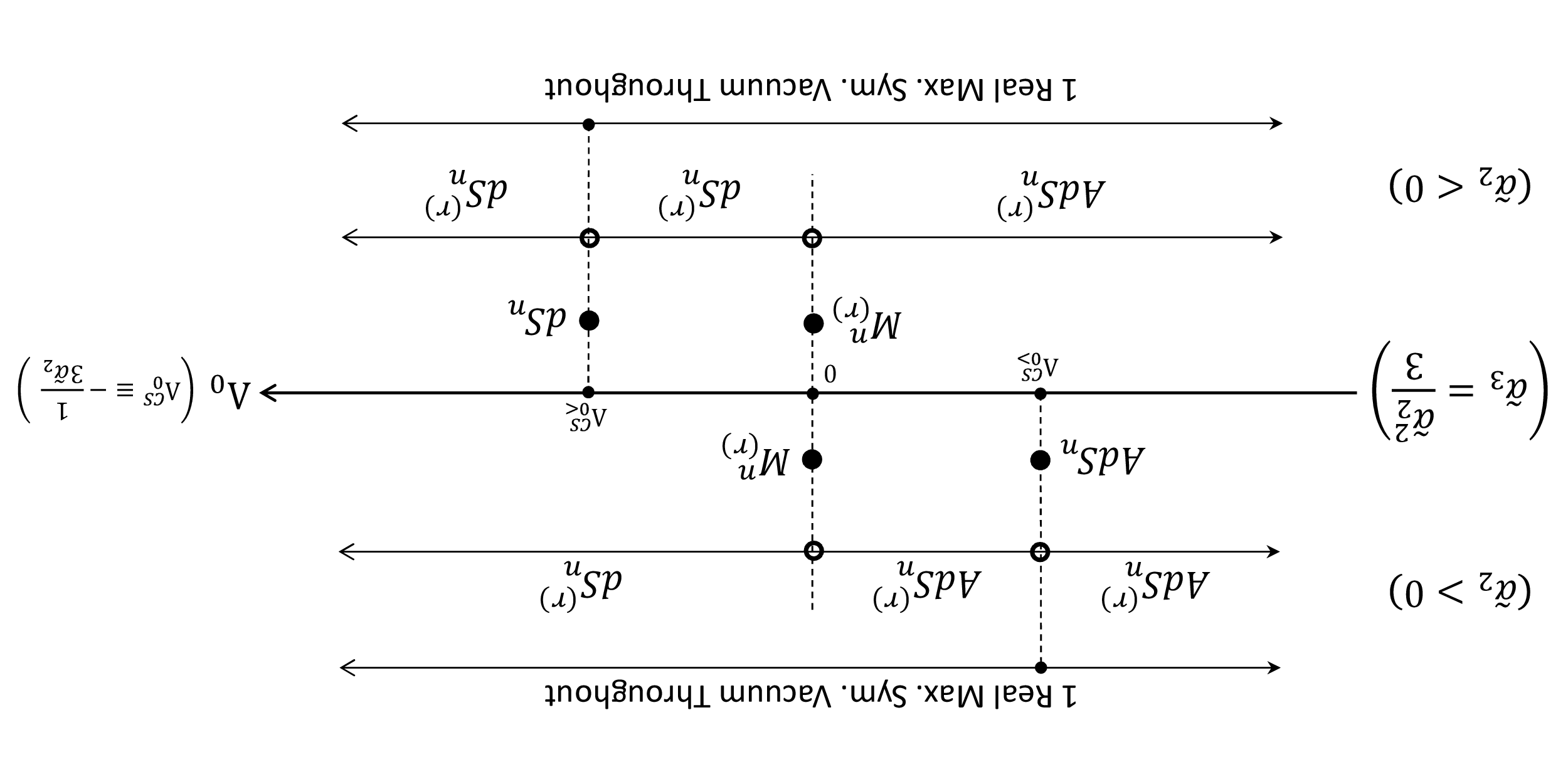}
\caption{\sl Maximally symmetric vacua of third-order Lovelock gravity in the special case $\tilde{\a}_3=\tilde{\a}_2^2/3$ in $n$ dimensions. The script \ql$(r)$" refers to the real solution (\ref{Rroot1}).}
\label{Fig3}
\end{figure}

\begin{figure}[t]
\centering
\includegraphics[width=3.75in,angle=270]{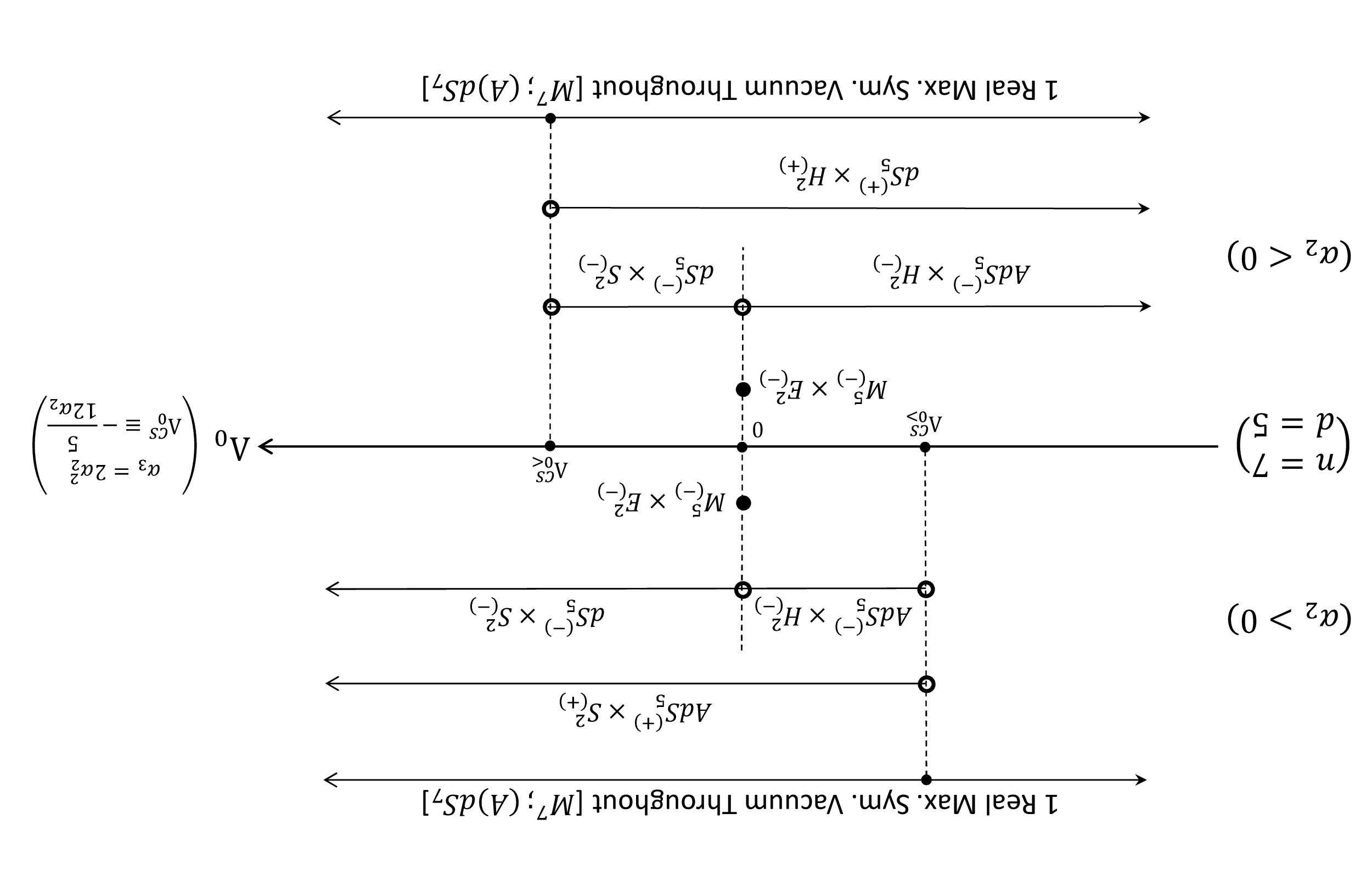}
\caption{\sl Direct product vacua of third-order Lovelock gravity when $n=7$ and $d=5$.}
\label{Fig16}
\end{figure}

The nature of the three roots (\ref{Rroot}) and (\ref{Croots}) depends on the sign of the quantity $\D_1^2-4\D_0^3$ in the following way\footnote{The first case is obvious, for it is the generic case. The second case can be understood as follows. When $\D_1^2-4\D_0^3=0$, there are three possibilities, which can be studied by considering $\D_1=0$, $\D_1>0$, and $\D_1<0$: If $\D_1=0$, then $\D_0=0$ and $A=0$. So in this case, (\ref{Rroot}) and (\ref{Croots}) seem indeterminate, but rewriting the equation (\ref{cubic}) as $$\l(\La+\frac{\tilde{\a}_2}{3\tilde{\a}_3}\r)^3-\frac{\D_0}{3\tilde{\a}_3^2}\l(\La+\frac{\tilde{\a}_2}{3\tilde{\a}_3}\r)+\frac{\D_1}{27\tilde{\a}_3^3}=0,$$ we see that in this case there is one triple real root. This is just the CS point. On the other hand, if $\D_1>0$ or $\D_1<0$, then $\D_1=2\D_0^{3/2}$ and $A=-\D_0^{1/2}$ or $\D_1=-2\D_0^{3/2}$ and $A=\D_0^{1/2}$, respectively. In either case, the complex part inside the square bracket in (\ref{Croots}) drops and we get three real roots two of which are always equal. And finally in the last case, $\D_1^2-4\D_0^3<0$, $A$ becomes complex and so $|A|^2=\D_0$. By using simple complex algebra, one can easily show that the roots (\ref{Rroot}) and (\ref{Croots}) are indeed real.}
\be\label{cases}
\begin{array}{ll}
1.&\tx{$\D_1^2-4\D_0^3>0$}\Rightarrow\tx{one real and two complex roots}, \\
2.&\tx{$\D_1^2-4\D_0^3=0$}\Rightarrow\tx{multiple real roots}, \\
3.&\tx{$\D_1^2-4\D_0^3<0$}\Rightarrow\tx{three distinct real roots}.
\end{array}
\ee
The CS point for the third order theory,  {\it i.e.} the point at which all three roots coincide, occurs when the quantities $\D_0= \D_1=0$.  The couplings at the CS points are then related according to 
\be\label{3rdCScon}
\tilde{\a}_3=\frac{\tilde{\a}_2^2}{3},\qquad \La_0=\La_0^{CS}\equiv-\frac{1}{3\tilde{\a}_2}.
\ee
The effective cosmological constant at the CS point is $\Lambda=-\frac{1}{\tilde{\a}_2}$
which can be either $dS_n$ for $\tilde{\a}_2<0$ or $AdS_n$ for $\tilde{\a}_2>0$.

It is difficult to investigate the full parameter space of third order Lovelock theory.
In order to make progress, we will restrict our attention to the two parameter family of theories satisfying $\D_0=0$, which will greatly simplify our analysis of product vacua, while still yielding interesting results.  We can now regard the third order coupling $\alpha_3$ as fixed in terms of $\alpha_2$ by the first condition in (\ref{3rdCScon}).
From (\ref{cases}) we see that with  $\Delta_0=0$ there will generically be one real root, which is found to be
\be\label{Rroot1}
\La_{1}=-\frac{1}{\tilde{\a}_2}\l[1-(1+3\tilde{\a}_2\La_0)^{1/3}\r].
\ee
which can be zero for $\La_0=0$, positive for $\La_0>0$, or negative for $\La_0<0$, and yielding respectively an $M^n$, $dS_n$, or $AdS_n$ vacuum. 
If in addition $\Delta_1=0$, which yields the second condition in (\ref{3rdCScon}), this is then the CS case with three coinciding real roots.
The corresponding maximally symmetric vacua are represented in \fig{Fig3}.  The vacua of Einstein gravity are recovered by taking the limit $\alpha_2\rightarrow 0$, which because 
$\Delta_0=0$ also takes the third order coupling to zero.

\begin{figure}[t]
\centering
\includegraphics[width=4.0in,angle=270]{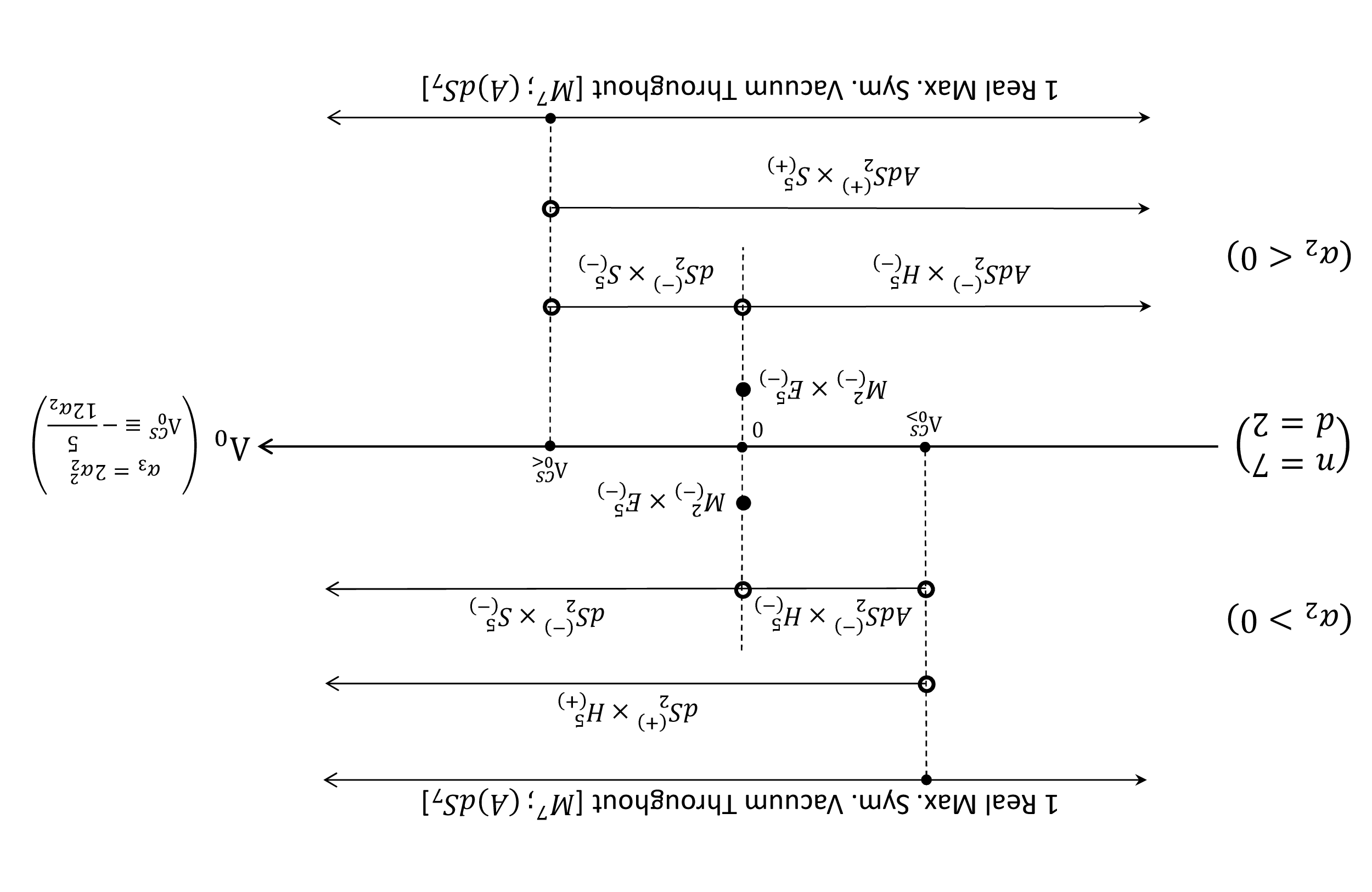}
\caption{\sl Direct product vacua of third-order Lovelock gravity when $n=7$ and $d=2$.}
\label{Fig13}
\end{figure}

We now turn to product vacua for third order Lovelock gravity.  The equations satisfied by the curvatures $K_1$ and $K_2$ are now given by
\begin{align}\label{third1}\nonumber
\alpha_3\left({(d-1)!\over (d-7)!}K_1^3 + {3(d-1)!D!\over (d-5)!(D-2)!} K_1^2K_2 \right.&\left.+ {3(d-1)!D!\over (d-3)!(D-4)!} K_1K_2^2  +{D!\over (D-6)!}K_2^3\right)\\ 
+\alpha_2\left( {(d-1)!\over (d-5)!}K_1^2 \right.&\left. + {2(d-1)!D!\over (d-3)!(D-2)!} K_1K_2 + {D!\over (D-4)!}K_2^2\right)\\ & +{(d-1)!\over (d-3)!}K_1+ {D!\over (D-2)!}K_2-2\Lambda_0 =0\nonumber
\end{align}
\begin{align}\label{third2} \nonumber
\alpha_3\left({(D-1)!\over (D-7)!}K_2^3 + {3(D-1)!d!\over (D-5)!(d-2)!} K_2^2K_1 \right.&\left.+ {3(D-1)!d!\over (D-3)!(d-4)!} K_2K_1^2  +{d!\over (d-6)!}K_1^3\right)\\
+\alpha_2\left( {(D-1)!\over (D-5)!}K_2^2 \right.&\left. + {2(D-1)!d!\over (D-3)!(d-2)!} K_2K_1 + {d!\over (d-4)!}K_1^2\right)\\ & +{(D-1)!\over (D-3)!}K_2+ {d!\over (d-2)!}K_1-2\Lambda_0 =0\nonumber
\end{align}
We will also restrict our analysis to $n=7$ dimensions, which is lowest dimension in which the third order Lovelock term is relevant.  We will consider in turn product vacua with 
$5+2$, $2+5$, $4+3$ and $3+4$ splits.

\begin{figure}[t]
\centering
\includegraphics[width=3.0in,angle=270]{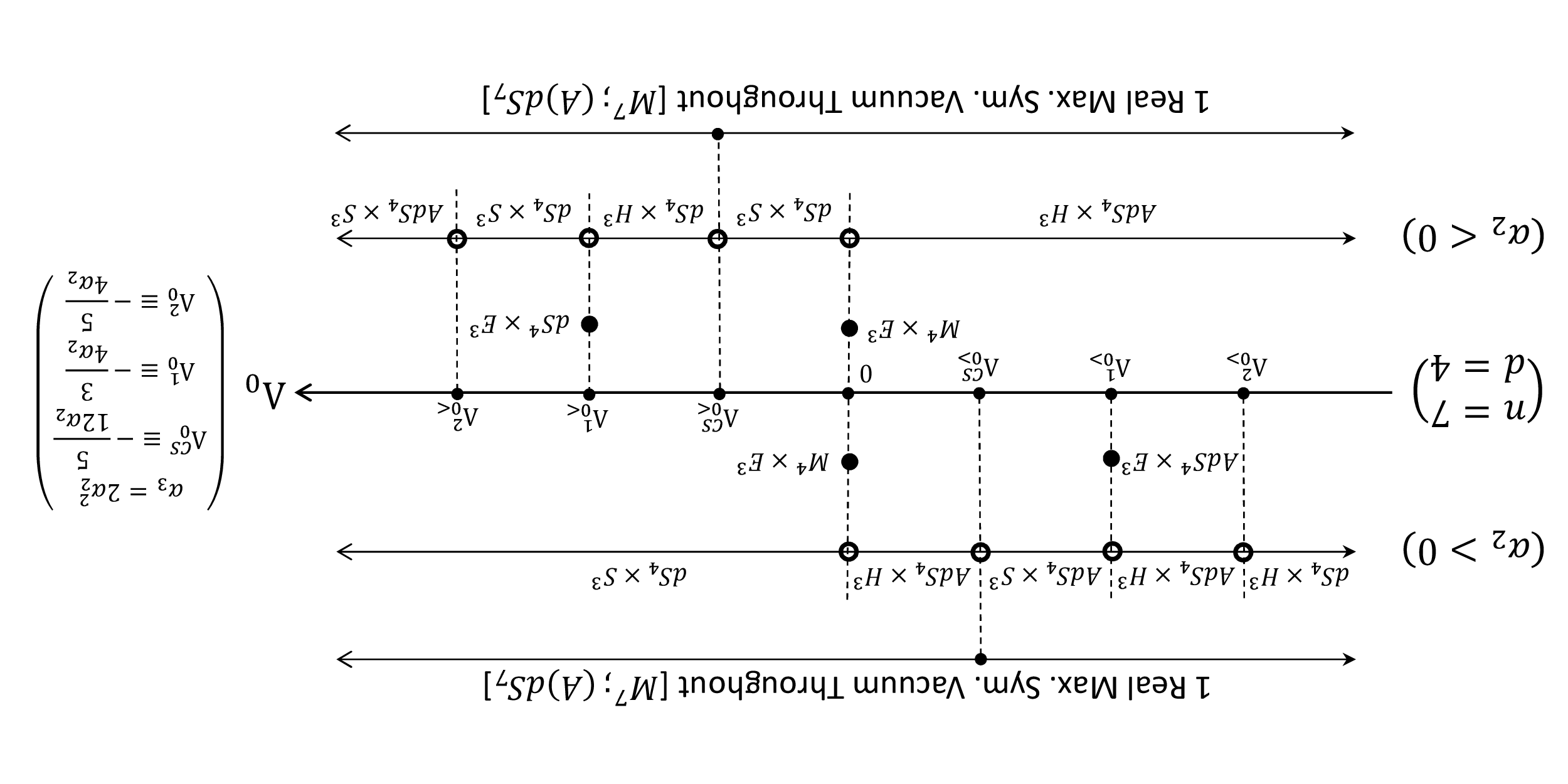}
\caption{\sl Direct product vacua of third-order Lovelock gravity when $n=7$ and $d=4$.}
\label{Fig15}
\end{figure}

\begin{figure}[t]
\centering
\includegraphics[width=3.0in,angle=270]{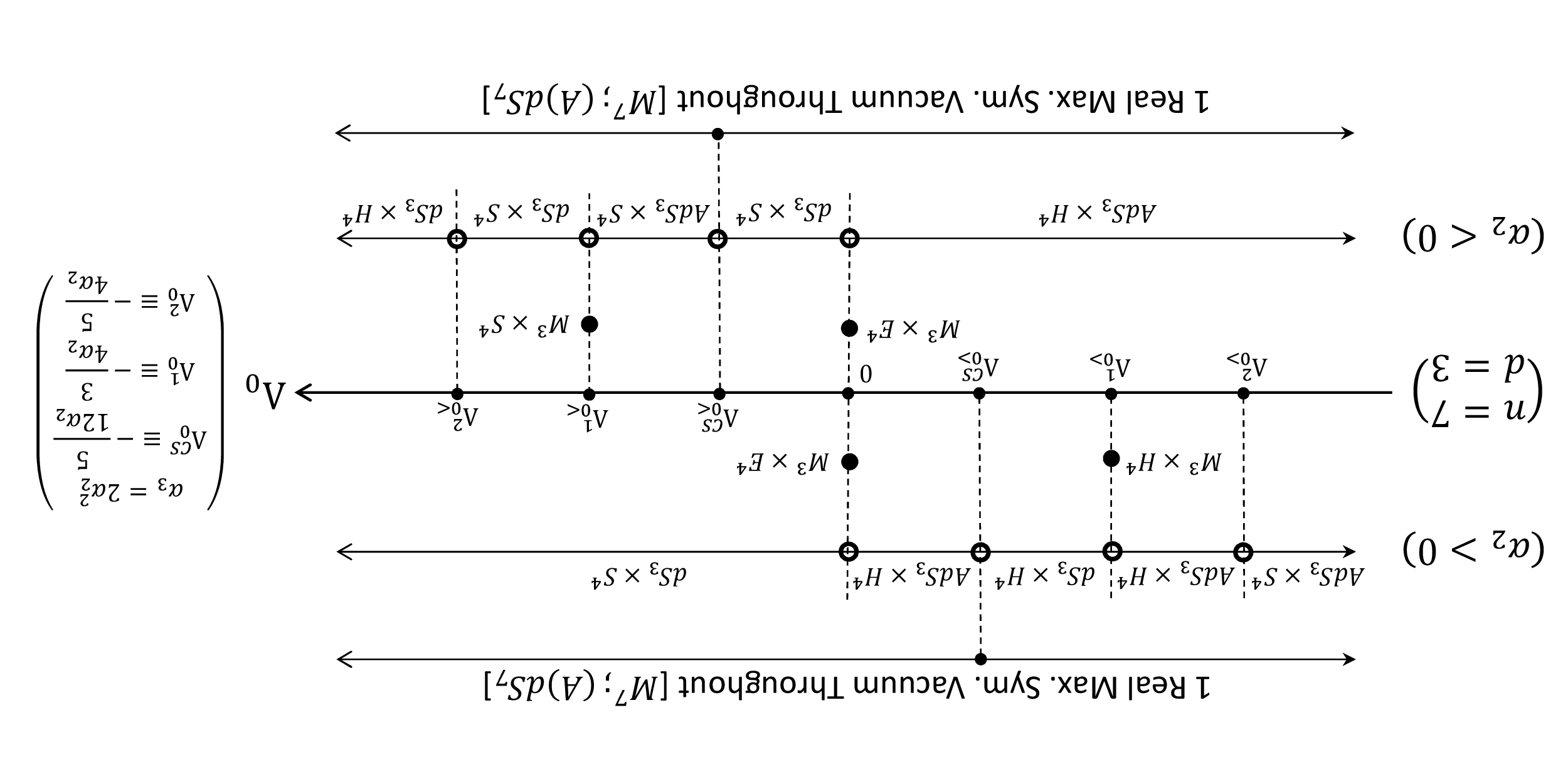}
\caption{\sl Direct product vacua of third-order Lovelock gravity when $n=7$ and $d=3$.}
\label{Fig14}
\end{figure}

Beginning with the $d=5$, $D=2$ split, we find that the equations of motion (\ref{third1}) and (\ref{third2}) reduce to 
\begin{align}
144\alpha_3 K_1^2K_2 +24\alpha_2 K_1^2 +48\alpha_2K_1K_2+12K_1+2K_2-2\Lambda_0 &=0\\
120\alpha_2 K_1^2 +20K_1-2\Lambda_0  &=0
\end{align}
We see that the second equation can be solved for $K_1$, with the first equation then determining $K_2$, giving
\be
K_1^\pm = {\Lambda_0^{CS}\over 5}C^\pm,\qquad K_2^\pm = -{4\Lambda_0^{CS}\over 5}{C^\pm\over C^\pm-1}
\ee
where $C^\pm=1\pm\sqrt{1-{\Lambda_0\over\,\,\,\,\,\Lambda_0^{CS}}}$, $\Lambda_0^{CS}=-{5\over 12\alpha_2}$ is the value of the cosmological constant at the CS point for the third order Lovelock theory in $n=7$ dimensions, and we have set 
$\alpha_3=2\alpha_2^2$ in accordance with our assumption that $\Delta_0=0$.  The character of these solutions and the range of $\Lambda_0$ covered is shown in Figure (\ref{Fig16}).  The case of a $2+5$ split reduces to the same set of equations, with the roles of the two curvatures $K_1$ and $K_2$ swapped.  The resulting solutions are displayed in Figure (\ref{Fig13}).

The structure of product vacua with a $d=4$, $D=3$ split is considerably more intricate.  In this case the equations of motion reduce to 
\begin{align}
72\alpha_2 K_1K_2+6K_1 +6K_2 -2\Lambda_0& =0\\
144\alpha_3 K_2K_1^2 +48\alpha_2K_1K_2+24\alpha_2K_1^2+2K_2+12K_1-2\Lambda_0 &=0.
\end{align}
After including the relation $\alpha_3=2\alpha_2^2$ this yields the curvatures
\be\label{3rdLGK1K2:n7d4}
K_1=\frac{2\La_0}{15(1-{\Lambda_0\over\,\Lambda_0^2})},\qquad
K_2=\frac{(1-{\Lambda_0\over\,\Lambda_0^1})\La_0}{5(1-{\Lambda_0\over\,\,\,\,\,\Lambda_0^{CS}})},
\ee
where the critical values of the cosmological constant at which one or the other of $K_1$ and $K_2$ change sign are given by 
$\Lambda_0^1=-{3\over 4\alpha_2}$, $\Lambda_0^2=-{5\over 4\alpha_2}$ and $\Lambda_0^{CS}=-{5\over 12\alpha_2}$.
These results are displayed in Figure (\ref{Fig15}).  The case of a $3+4$ split is simply obtained by swapping the curvatures $K_1$ and $K_2$ and is displayed in Figure (\ref{Fig14}).

\begin{figure}[t]
\centering
\includegraphics[width=3.0in,angle=270]{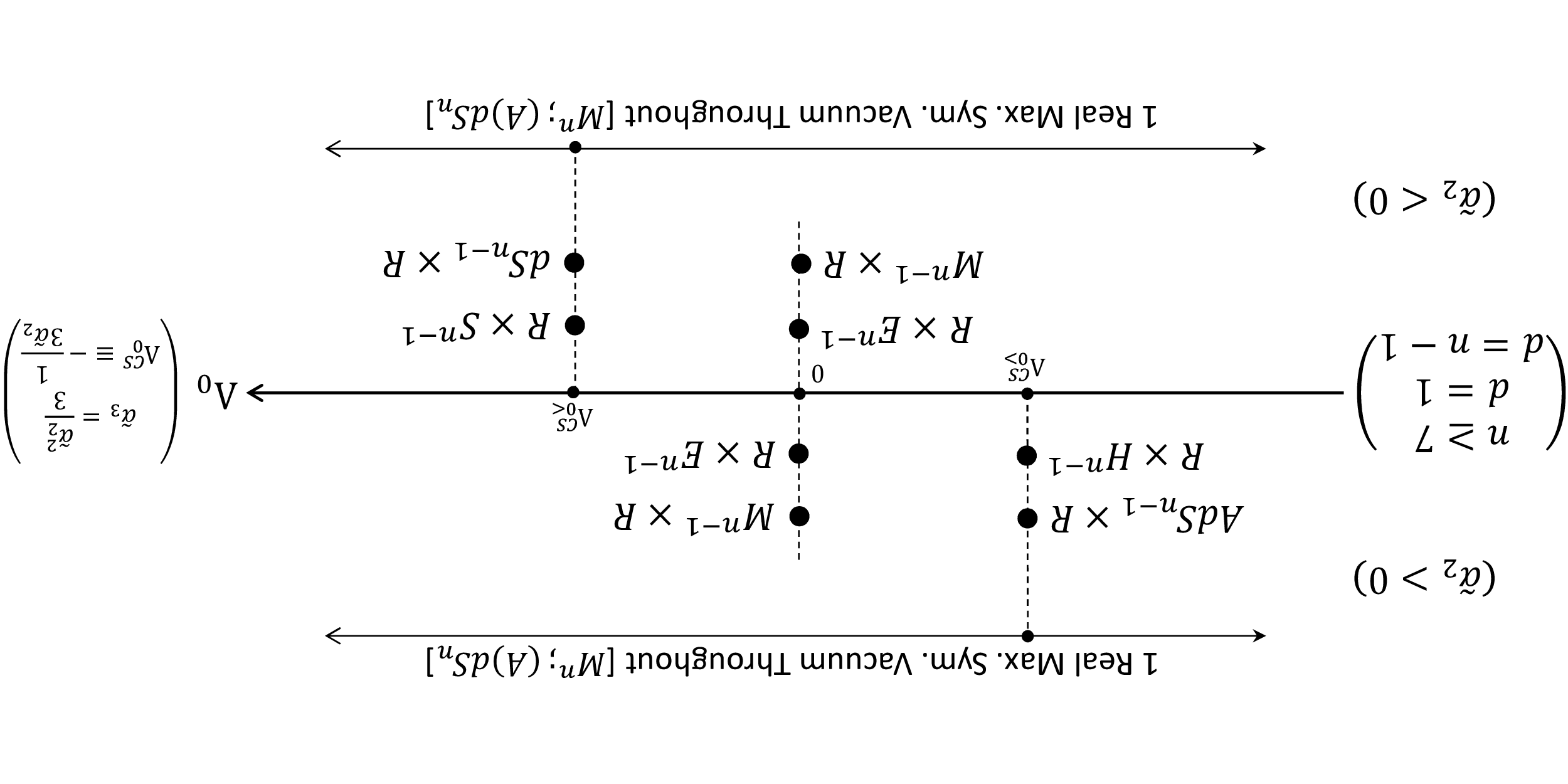}
\caption{\sl Direct product vacua of third-order Lovelock gravity in the case of one-dimensional submanifolds in $n$ dimensions.}
\label{Fig12}
\end{figure}

Finally, we consider products with one dimensional factors, $d=n-1$ and $d=1$.  Taking $d=1$, we find that as in the Gauss-Bonnet case 
solutions exist only for $\Lambda_0=0$ with $K_2=0$ and for $\Lambda_0=-{1\over 3\tilde\alpha_2}\equiv\Lambda_0^{CS}$ with $K_2={6\over (n-1)(n-2)}\Lambda_0^{CS}$.  
This result\footnote{The solutions with $\Lambda_0=0$ are actually present for general values of $\alpha_2$ and $\alpha_3$.}  
along with the corresponding result for $d=n-1$ is displayed in Figure (\ref{Fig12}).   It is intriguing that the CS value of the couplings show up again here.  It would be interesting to understand the case of product vacua with one dimensional factors more generally.

\section{Conclusion}\label{conclude}

Einstein gravity has maximally symmetric vacua for all values of the cosmological constant and in all spacetime dimensions.  However, Lovelock theories can have symmetry breaking regions of coupling space,  in which no maximally symmetric vacua exist.  We have carried out a partial survey of alternative, reduced symmetry vacua in Lovelock theories that are products of lower dimensional maximally symmetric space(times), with particular interest in whether such vacua exist cover symmetry breaking regions of coupling space.  Our findings on this question show indications of interesting structure.
Gauss-Bonnet gravity in any dimension has such a symmetry breaking region of coupling space.  We looked at product vacua in $n=5$ and $n=6$ dimensions, finding sharply different results.  While product vacua cover the entire symmetry breaking region of coupling space in $n=5$ dimensions, in $n=6$ dimensions such vacua only cover a small portion of the symmetry breaking area.  

In $n=7$ dimensions, the third order Lovelock interaction becomes physically relevant, and so long as its coupling is nonzero at least one maximally symmetric vacuum will exist.  We have looked at product vacua in this theory, restricting our focus to a tractable region of coupling space.  We found that  $5+2$ and $2+5$ dimensional products again exist throughout this region.  However, $4+3$ and $3+4$ dimensional products exist only over a portion of coupling space. 

It would be interesting to extend this study further.  For example, one could look at  product vacua in Gauss-Bonnet gravity beyond $n=6$ dimensions, {\it i.e.} setting the couplings of the relevant higher order Lovelock terms to zero.  Our results in $n=5,6$ dimensions would be consistent with a number of possible patterns.  For example, it might turn out that product vacua cover all of the symmetry breaking region of coupling space only in $n=5$ dimensions.  Alternatively, it could be that there is an alternation between even and odd dimensions, with product vacua covering the symmetry breaking region fully  in odd dimensions.  As the dimension gets higher, it might also be natural to consider product vacua with more than two factors.  It would be particularly interesting to map out the product vacua of $4$th order Lovelock theory, which will also have a symmetry breaking regime, in $n=9,10$ dimensions, where it includes all the relevant Lovelock terms.  However, even if one considers only a particular subspace of the full set of couplings, the equations for the curvatures will be quartic and difficult to analyze.  It is also important to note that not all maximally symmetric vacua in Lovelock gravity are stable.  For example, the Gauss-Bonnet branch of vacua in Gauss-Bonnet gravity suffers from a ghost instability \cite{Boulware:1985wk}.  As noted in \cite{Canfora:2008iu}, it will be important to study the stability of product solutions such as those found here, in order to determine the true vacua of the theory.

Finally, it would be interesting to consider the potential physical relevance of transitions across critical surfaces in coupling space in which the number of maximally symmetric vacua change.  This could happen, for example, if the cosmological constant were dynamical\footnote{There has recently been a good deal of interest in considering the cosmological constant as a thermodynamic variable in the context of black hole physics (see {\it e.g.} \cite{Kastor:2009wy,Cvetic:2010jb,Kubiznak:2012wp} and references thereto).  The thermodynamics of varying Lovelock couplings has also been considered in \cite{Kastor:2010gq}.}.  If the cosmological constant crossed into the symmetry breaking region of coupling space in $n=5$ dimensional Gauss-Bonnet gravity, it might be possible to transition from an $AdS_5$ vacuum to an $AdS_3\times S^2$ vacuum as in the top half of  Figure (\ref{Fig8}). 



\subsection*{Acknowledgements}

\c{C}. \c{S}. wishes to thank ACFI for hospitality and also thanks the Scientific and Technological Research Council of Turkey (T{\"U}B\.{I}TAK) for financial support under the Programme BIDEB-2219.

\end{document}